\documentclass
[pra,onecolumn,notitlepage,nofootinbib,superscriptaddress,showkeys]{revtex4-1}%
\usepackage[T1]{fontenc}
\usepackage[latin9]{inputenc}
\usepackage{amsmath}
\usepackage{amssymb}
\usepackage[brazil,english]{babel}
\usepackage{graphicx}
\usepackage[usenames,dvipsnames]{color}
\usepackage{longtable}
\usepackage{dcolumn}
\usepackage{ulem}
\usepackage{amsfonts}
\usepackage{hyperref}%
\usepackage{xcolor}
\providecommand{\U}[1]{\protect\rule{.1in}{.1in}}
\begin{document}

\title{Weyl conjecture and thermal radiation of finite systems}

\author{M. C. Baldiotti}
\email{baldiotti@uel.br}
\author{M. A. Jaraba}
\email{jaraba.marcosrod@uel.br}
\affiliation{Departamento de F\'{\i}sica, Universidade Estadual de Londrina, CEP 86051-990, Londrina-PR, Brazil.}
\author{L. F. Santos}
\email{luisf@usp.br}
\affiliation{Instituto de F\'{\i}sica, Universidade de S\~{a}o Paulo, Caixa Postal 66318, CEP 05315-970, S\~{a}o Paulo-SP, Brazil.}
\author{C. Molina}
\email{cmolina@usp.br}
\affiliation{Escola de Artes, Ci\^{e}ncias e Humanidades, Universidade de S\~{a}o Paulo, Avenida Arlindo Bettio 1000, CEP 03828-000, S\~{a}o Paulo-SP, Brazil.}

\begin{abstract}
In this work, corrections for the Weyl law and Weyl conjecture in $d$
dimensions are obtained and effects related to the polarization and area term
are analyzed. The derived formalism is applied on the quasithermodynamics of
the electromagnetic field in a finite $d$-dimensional box within a semi-classical treatment. In this context,
corrections to the Stefan-Boltzmann law are obtained. Special attention is
given to the two-dimensional scenario, since it can be used in the
characterization of experimental setups. Another application concerns acoustic perturbations in a quasithermodynamic generalization of Debye model for a finite solid in
$d$ dimensions. Extensions and corrections for known results and usual
formulas, such as the Debye frequency and Dulong-Petit law, are calculated.
\newline


\end{abstract}

\maketitle

\section{Introduction}

The analysis of thermal radiation is widespread in a large variety of
finite-temperature systems. Theoretical research includes treatments based on
fluid dynamics and/or quantum field theory. Laboratory and observational
applications involve particle phenomenology in colliders, properties of solids
in laboratories, characteristics of the cosmic microwave background in
dedicated observatories, among many more setups. From lower-dimensional
settings to models with arbitrary number of dimensions, thermal radiation
frequently plays a significant role.

A common procedure for the investigation of thermal phenomena is based on the
definition of a suitable thermodynamic limit within a statistical mechanics
model. Usually, the transition from statistical mechanics to thermodynamics
implies that the boundary conditions are neglected. In this way, the
dependence of the obtained results with the actual volume and shape of the
physical system is not considered. Although this approach is useful for many
purposes, the strict thermodynamic limit disregards many interesting insights
about the actual system of interest. One way to mitigate this problem is to
define an intermediary regime between the pure statistical-mechanic treatment
and the strict thermodynamic regime, where the volume of the system is large,
but finite. This is the so-called quasithermodynamic limit \cite{Maslov1}.

On a very general level, the characteristics of thermal radiation in a given
setup are directly linked to the asymptotic distribution of eigenvalues of a
suitable wave equation. One of the first investigators to explore this
connection was Rayleigh, studying the problem of stationary acoustic waves in
a cubic room \cite{Ray, Ray2}. Rayleigh's analysis showed the importance of a
term proportional to the volume of the room and to the cube of the mode
frequency (the $V\cdot\nu^{3}$ term). This result appeared in the (incorrect)
description of the thermal radiation with the Rayleigh-Jeans law. Eventually
Planck improved this purely classical analysis, but even within the quantum
description the eigenvalue distribution remains unaltered. The same problem,
and hence with the same $V\cdot\nu^{3}$ term as the result, emerges in the
investigation of the vibration modes in a solid, with the so-called Debye
model. In this treatment, Debye proposed that the asymptotic behavior of the
eigenvalues do not depend on the shape of the solid. This proposal was
rigorously proved by Weyl, and today it is known as the Weyl law. An overview
of this development can be seen in \cite{Arendt,Ivrii2}.

A central question considering the strict thermodynamic regime is when the
approximation of infinite volume adequately describes a real physical system.
For the treatment of this issue, it is necessary an estimate of the terms that
are being dropped in the thermodynamic limit. A first step in this direction
is given by Weyl conjecture, which predicts corrections proportional to the
area of the body. This conjecture has been proven in a variety of domains and in
 this process, several methods can be applied. For instance, asymptotically
expanding the solutions of the Helmholtz equation, using the Neumann-Poincar\'{e}
construction for the Brownell Green's function and considering the
decomposition of the mode density via multiple reflection expansion \cite{Balian, BalBloc1}.

For many important problems, involving for example the electromagnetic field,
vector solutions and polarization effects must be considered. For this
purpose, a possible approach is to decompose the vector fields into solutions
of the suitable scalar wave equation \cite{BalBloc2,Balian,Baltes Baltes,Liu}.
Following this program, many works in the pertinent literature indicate that
the area term (i.e., the term of Weyl conjecture) does not participate in
describing the behavior of the thermal electromagnetic radiation.

Previous comments refer mainly to usual scenarios with three dimensions. But
the relevance of the proposed treatment appears in models with different
dimensionalities. For instance, the thermodynamic and quasithermodynamic
analyses of two-dimensional systems have practical applications in the
so-called single-layer materials \cite{A15}, with highlights to the graphene
\cite{B1}. As a more theoretical application, we can mention the thermodynamic
properties of photon spheres \cite{photon-spheres}, thermal radiation of the
two-dimensional bosonic and fermionic modes of black holes \cite{Geoffrey}, as
well the description of thermodynamics properties of BTZ black holes \cite{btz
bh}. Considering three-dimensional systems, corrections in thermal radiation
play an important role in the analysis of the background microwave radiation
\cite{Garcia-Garcia}. The development is also relevant in the thermodynamic
description of the phenomenon of sonoluminescence (hot spot theory), in which
pulses of light are created by means of the insertion of sound waves into
liquids or gases \cite{lumin2}. Systems with more than three spatial
dimensions are also explored. For instance, Hawking emission of a black hole
is altered in brane-world scenarios
\cite{Cardoso:2005mh,Abdalla:2006qj,rge:2014kra}. Thermal radiation phenomena
associated with gravitational emission in brane-world models could be
significant in the early Universe. Similar effects could be expected in
colliders and near active astrophysical objects, within models involving extra
dimensions (see for example \cite{Maartens:2010ar} and references therein).
Proposals linking anti-de Sitter geometries and conformal field theories
(AdS/CFT correspondences) offer a great variety of applications for
$d$-dimensional thermodynamic results
\cite{Baldiotti:2016lcf,Baldiotti:2017ywq,Elias:2018yct,Fontana:2018drk}.

In the present work, Weyl law and its extensions are explored through an
intuitive approach. Quasithermodynamic analysis and its description of systems
with a finite volume and relevant boundary conditions are a central issue in
this article. The developed formalism is applied on the quasithermodynamics of
the electromagnetic field in a cavity and acoustic perturbations in a solid. Generalizations
for $d$-dimensional setups are derived. One of the contributions of this work
is to incorporate polarization effects directly into the asymptotic expansions
of the mode distributions, using mixed boundary conditions. We emphasize the
role of the area term, showing that, under certain conditions, a distinct
quasithermodynamic behavior emerges.

In addition to the correction coming from the eigenvalue distribution, an
expected characteristic of the phenomenology of black-body radiation in finite
cavities is the existence of a minimum energy.\footnote{This lower bound on
the energy of the system would be associated to the quantum vacuum, according
to \cite{stefan 2d}.} For the two-dimensional case, this issue was studied in
\cite{stefan 2d}. Generalizing this development for arbitrary dimensions, our
treatment allows us to compare the effects associated to the minimum energy
with those related to corrections of the spectral distribution.

This work is organized as follows. In section~\ref{cce}, the proposed
formalism associated to the Weyl law is established. In section~\ref{beyond},
with the techniques introduced, higher-order corrections to the Weyl law are
obtained. More complex setups are considered in section~\ref{comment}, where
mixed boundary conditions and degeneracies are treated. In
section~\ref{thermodynamics} we turn to physics applications, linking the
results previously obtained with thermodynamics and quasithermodynamics. In
section~\ref{photons}, the thermodynamic treatment of the electromagnetic
field in a finite cavity is conducted. The quasithermodynamics of acoustic perturbations is
considered in section~\ref{phonons}, where Debye model is analyzed and
extended. In section~\ref{conclusion} final comments and future perspectives
are presented. Further details on the calculation of the hypervolume
associated to the axes and counting functions are discussed in
appendices~\ref{eixos} and \ref{pascal}.

\section{\label{cce} Scalar field and Weyl law}

Electromagnetic and mechanic perturbations in cavities and solids are the main
interest in the present work. However, as we will see later on, the
thermodynamics of those systems can be described in terms of a simpler scalar
perturbation. Let us consider a scalar field $\psi\left(  x_{1},\ldots
,x_{d}\right)  $ in $d$ dimensions, confined in a cubic cavity of size $L$,
which respects the Helmholtz equation,
\begin{equation}
\nabla_{d}^{2}\psi\left(  x_{1},\ldots,x_{d}\right)  +k^{2}\psi\left(
x_{1},\ldots,x_{d}\right)  =0\,,\ x_{i}\in\Omega_{d} \, ,\ k\in\mathbf{%
\mathbb{R}
}\,.\label{helmholtz d dimensoes}%
\end{equation}
In Eq.~\eqref{helmholtz d dimensoes}, $\nabla_{d}^{2}$ denotes the
$d$-dimensional Laplacian. The hypervolumes of domain $\Omega_{d}$ and its
boundary $\partial\Omega_{d}$ are given respectively by
\begin{equation}
\left\vert \Omega_{d}\right\vert =L^{d}\,,\,\,\,\left\vert \partial\Omega
_{d}\right\vert =2dL^{d-1}\,.\label{volume do dominio dD}%
\end{equation}
Two different boundary conditions for Eq.~\eqref{helmholtz d dimensoes} will
be initially explored, namely the Neumann condition ($\partial_{n}\psi=0$ at
$\partial\Omega_{d}$) and the Dirichlet condition ($\psi=0$ at $\partial
\Omega_{d}$).

Solutions of the $d$-dimensional Helmholtz equation~\eqref{helmholtz d
dimensoes} can be constructed using the one-dimensional version of
\eqref{helmholtz d dimensoes}, which are
\begin{equation}
\phi_{n}^{\pm}\left(  x\right)  =\frac{e^{-i\pi/4}\pm e^{i\pi/4}}{\sqrt{4L}%
}\left(  e^{ik_{n}x}\pm e^{-ik_{n}x}\right)  \, ,\ k_{n}=\frac{\pi}{L}n \, ,
\,\, n=\left\{
\begin{array}
[c]{c}%
0,1,2,\ldots\left(  +\right) \\
1,2,3,\ldots\left(  -\right)
\end{array}
\right.  \,.
\end{equation}
The label ($+$) indicates solutions satisfying Neumann conditions, while ($-$)
denotes solutions obeying Dirichlet conditions. From the one-dimensional
solutions, $d$-dimensional generalizations can be constructed as
\begin{equation}
\psi^{\pm}\left(  x_{1},\ldots,x_{d}\right)  =\phi_{n_{1}}^{\pm}(x_{1}%
)\phi_{n_{2}}^{\pm}(x_{2})\cdots\phi_{n_{d}}^{\pm}(x_{d}) \, ,
\end{equation}
with
\begin{equation}
k^{2}=\frac{\pi^{2}}{L^{2}}(n_{1}^{2}+\cdots+n_{d}^{2})~.
\label{definicoes uteis}%
\end{equation}
It should be noted that, for the Dirichlet case, the solutions with $n_{i}=0$
should be excluded in order to avoid the null solution.

Considering Eq.~\eqref{definicoes uteis}, the set of solutions of the
Helmholtz equation~\eqref{helmholtz d dimensoes} can be labeled by points in a
discrete lattice. Specifically, we consider the $d$-dimensional Cartesian
lattice $L_{\epsilon}^{d}$,
\begin{equation}
L_{\epsilon}^{d}\equiv\left\{  (\epsilon_{1},\epsilon_{2},\ldots,\epsilon
_{d})\in\mathbb{R}^{d}:\epsilon_{i}=\epsilon n_{i} \, ,\ n_{i}=0,1,2,\ldots
\right\}  \,,
\end{equation}
where the (real and positive) $\epsilon$ is the lattice parameter. We denote
the region $\tilde{\Omega}_{d}\subset\mathbb{R}^{d}$ as one of the $2^{d}$
sections of the $d$-dimensional sphere of radius $k$, i.e.,
\begin{equation}
\tilde{\Omega}_{d}\equiv\left\{  (\epsilon_{1},\epsilon_{2},\ldots
,\epsilon_{d})\in\mathbb{R}^{d}:\epsilon_{1}^{2}+\epsilon_{2}^{2}%
+\cdots+\epsilon_{d}^{2}<k^{2} \, ,\ \epsilon_{i}\geqslant0\right\}  \,.
\end{equation}

The main questions addressed in the present work can be formulated as counting
problems. In this direction, let us define a counting function $N^{(d)}(k)$
as
\begin{equation}
N^{\left(  d\right)  }\left(  k\right)  =\left\{  \#\left(  n_{1},\ldots
,n_{d}\right)  :\frac{\pi^{2}}{L^{2}}\left(  n_{1}^{2}+\cdots+n_{d}%
^{2}\right)  <k^{2}\right\}  \,. \label{eq7}%
\end{equation}
The function $N^{(d)}(k)$ can be expressed as
\begin{equation}
N^{\left(  d\right)  }(k)=\mathrm{card}\left(  L_{\epsilon}^{d}\cap
\tilde{\Omega}_{d}\right)  \, , \,\, \epsilon=\frac{\pi}{L}\,,
\label{funcontagem}%
\end{equation}
where \textquotedblleft$\mathrm{card}$\textquotedblright\ denotes the
cardinality of the set \cite{Suppes}. It should be noted that the complete
dependence of $N^{\left(  d\right)  }(k)$ on $k$ is described by
$\tilde{\Omega}_{d}$, since the lattice is (for now) fixed. We are interested
in the asymptotic behavior of $N^{(d)}(k)$ as $k\rightarrow\infty$.

One method for the analysis is to adopt a \textquotedblleft coarse
grained\textquotedblright\ version of the lattice, where the counting of the
discrete points is substituted by the calculation of volumes. More
specifically, the function $N^{(d)}(k)$ is approximated by the volume
$V_{d}(k)$ generated by hypercubes of side length $\epsilon$, centered on the
points of the lattice belonging to 
$L_{\epsilon}^{d}\cap\tilde{\Omega}_{d}(k)$. 
We illustrate this coarse graining in Figs.~\ref{fig1} and \ref{fig2} of the next section. The volume of each hypercube is $\epsilon^{d}$, and hence
$V_{d}\left(  k\right)  =\epsilon^{d}N^{(d)}(k)$. The total volume
$V_{d}\left(  k\right)  $ tends to infinity as $k\rightarrow\infty$,
$N^{(d)}(k)\rightarrow\infty$ and $\epsilon$ is kept fixed.

Expression~\eqref{funcontagem} and the coarse graining introduced are a
well-known algorithm for the counting process, in which the lattice is kept
fixed. In the present work, we propose an alternative approach. Instead of
fixing $\epsilon$ (that is, the lattice), we fix the volume $V_{d}\left(
k\right)  $. In other words, we propose to maintain the domain fixed and
adjust the lattice. This can be accomplished rescaling $\epsilon_{i}$,
\begin{equation}
\epsilon_{i}\longrightarrow\epsilon_{i}=k^{-1}\epsilon_{i}=\epsilon n_{i} \, ,
\,\, \epsilon=\frac{\pi}{kL}\,.
\end{equation}
With the rescaling, both the counting function $N^{\left(  d\right)  }\left(
k\right)  $ and the volume $V_{d}$ become dependent only on the lattice, that
is,
\begin{equation}
V_{d}\left(  \epsilon\right)  =\epsilon^{d}N^{(d)}(\epsilon)\,.
\label{omega-d}%
\end{equation}
Intuitively, $V_{d}(\epsilon)$ should tend to the volume $|\tilde{\Omega}%
_{d}|$ as the lattice becomes more dense ($\epsilon\rightarrow0^{+}$), where
\begin{equation}
\left\vert \tilde{\Omega}_{d}\right\vert =2^{-d}\omega_{d} \, ,\ \omega_{d}%
=\frac{\pi^{d/2}}{\Gamma\left(  \frac{d}{2}+1\right)  }\,,
\end{equation}
and $\Gamma$ denotes the usual gamma function. The precise development of this
relation will lead to the so-called Weyl law, which will be explicitly shown
with the approach proposed in this work.

To clarify the notation, let us denote the counting function for the Neumann
case as $N_{+}^{(d)}$, and analogously $N_{-}^{(d)}$ for the Dirichlet case.
We focus on the difference in the counting problem considering the Neumann and
Dirichlet setup. This difference is a result of the inclusion of the points
belonging to the $\epsilon_{i}$-axis for the calculation of $N_{+}^{(d)}$, and
the exclusion for $N_{-}^{(d)}$.

Let us treat the Neumann boundary condition first, since (as will be shown in
later sections) $N_{-}^{(d)}$ can be expressed in terms of $N_{+}^{(d)}$. The
counting function $N_{+}^{(d)}$ can be written as
\begin{equation}
N_{+}^{(d)}(\epsilon)=\sum_{n_{1}=0}^{M(\epsilon)}\cdots\sum_{n_{d-1}%
=0}^{M(\epsilon)}\left\lfloor \frac{\sqrt{1-\epsilon^{2}(n_{1}^{2}%
+\cdots+n_{d-1}^{2})}}{\epsilon}\right\rfloor \,,\ M(\epsilon)=\left\lfloor
\frac{1}{\epsilon}\right\rfloor \,, \label{eqfuncon}%
\end{equation}
with $\lfloor x\rfloor$ representing the integer function (or floor function)
of $x$ \cite{Graham}. Using the sawtooth function $\eta(x)$, $\eta(x)\equiv
x-\lfloor x\rfloor$, Eq.~\eqref{omega-d} is written as
\begin{equation}
V_{d}^{+}(\epsilon)=\epsilon^{d-1}\sum_{n_{1}=0}^{M(\epsilon)}\cdots
\sum_{n_{d-1}=0}^{M(\epsilon)}\sqrt{1-\epsilon^{2}(n_{1}^{2}+\cdots
+n_{d-1}^{2})}-\epsilon^{d}\sum_{n_{1}=0}^{M(\epsilon)}\cdots\sum_{n_{d-1}%
=0}^{M(\epsilon)}\eta\left(  \frac{1}{\epsilon}\right)  \,. \label{eqFG}%
\end{equation}
Since $0\leq\eta<1$, the second term of Eq.~\eqref{eqFG} obeys the following
property:
\begin{equation}
0\leq\epsilon^{d}\sum_{n_{1}=0}^{M\left(  \epsilon\right)  }\cdots
\sum_{n_{d-1}=0}^{M\left(  \epsilon\right)  }\eta<\epsilon\left[
\epsilon+\epsilon M\left(  \epsilon\right)  \right]  ^{d-1}\,,
\end{equation}
implying that
\begin{equation}
\lim_{\epsilon\rightarrow0^{+}}{\epsilon^{d}\sum_{n_{1}=0}^{M\left(
\epsilon\right)  }\cdots\sum_{n_{d-1}=0}^{M\left(  \epsilon\right)  }\eta
}\left(  \frac{1}{\epsilon}\right)  =0\,. \label{eqempar}%
\end{equation}

Previous results lead to the Weyl law, as we will show in the following.
Considering the first term in Eq.~\eqref{eqFG}, we observe that%
\begin{equation}
\epsilon=\frac{M^{-1}}{1+\eta M^{-1}}\Rightarrow\lim_{\epsilon\rightarrow
0^{+}}{\epsilon}=\lim_{M\rightarrow\infty}\frac{M^{-1}}{1+\eta M^{-1}}%
=\lim_{M\rightarrow\infty}M^{-1} \, ,
\end{equation}
and we can write
\begin{align}
&  \lim_{\epsilon\rightarrow0^{+}}{\epsilon^{d-1}\sum_{n_{1}=0}^{M\left(
\epsilon\right)  }\cdots\sum_{n_{d-1}=0}^{M\left(  \epsilon\right)  }%
\sqrt{1-\epsilon^{2}\left(  n_{1}^{2}+\cdots+n_{d-1}^{2}\right)  }}\nonumber\\
&  =\lim_{M\rightarrow\infty}{\ \frac{1}{M^{d-1}}\,\sum_{n_{1}=0}^{M\left(
\epsilon\right)  }\cdots\sum_{n_{d-1}=0}^{M\left(  \epsilon\right)  }%
\sqrt{1-\frac{n_{1}^{2}+\cdots+n_{d-1}^{2}}{M^{2}}}}\,. \label{soma_riemann}%
\end{align}
The last terms in Eq.~\eqref{soma_riemann} is the Riemann sum which gives the
volume of $\tilde{\Omega}_{d}$. Hence, using Eq.~\eqref{eqempar}, we conclude
that
\begin{equation}
\lim_{\epsilon\rightarrow0^{+}}{V_{d}^{+}}\left(  {\epsilon}\right)
=2^{-d}\omega_{d}\,. \label{weyl-law}%
\end{equation}
Result~\eqref{weyl-law} is the Weyl law for the scalar field in a cubic
cavity, with Neumann boundary conditions.

Although the Weyl law is a well-known result, the procedure adopted here (that
is, a coarse graining with a scaling which fixes the domain and modifies the
lattice) can be employed in the improvement of the result~\eqref{weyl-law}.
Also, this approach will allow us to consider different boundary conditions
and polarization effects.

\section{\label{beyond} Beyond the Weyl conjecture}

Weyl law can be considered as the zero-order correction of the counting
function. The first-order correction was conjectured by Weyl and later proven
by Ivrii \cite{Ivrii1}. In our notation, the Weyl conjecture can be written
as
\begin{equation}
\epsilon^{d}N_{\pm}^{(d)}(\epsilon)=2^{-d}\omega_{d}\pm2^{-d}d\omega
_{d-1}\epsilon+O(\epsilon^{w_{d}}) \, ,\label{eqNDe}%
\end{equation}
where $w_{d}=\left(  d^{2}-d+1+1/d\right)  /\left(  d-1\right)  $
\cite{Brownell}. In the \textquotedblleft big-$O$\textquotedblright\ sense,
no higher-order corrections are known\footnote{One writes $f(x)=O(g(x))$ if
there exists a real $\delta>0$ and $x_{0}$ such that $\left\vert f\left(
x\right)  \right\vert \leq\delta g\left(  x\right)  $ for all $x>x_{0}$. That
is, $f$ is smaller than $g$ as $x\rightarrow\infty$, and the asymptotic
behavior of $f$ is bounded by the function $g$.} \cite{Ivrii1, Brownell}.
Although for the scalar case the above correction is always the most relevant,
as we will see, when considering the vector problem this correction may
cancels. Therefore, its important to go beyond the first correction. For this
goal a special type of \textquotedblleft averaged
corrections\textquotedblright\ can be considered. It is defined using Gaussian
logarithmic averages, or the called Brownell's $\tilde{O}$ formalism
\cite{Brownell}. One says that $f=\tilde{O}\left(  g\right)  $ if, for some
$x_{0}$,
\begin{equation}
\left\vert \int_{x_{0}}^{\infty}e^{-\frac{1}{2}\rho^{2}\left(  \ln\frac
{y}{x^{\prime}}\right)  ^{2}}~\left.  \frac{df(x)}{dx}\right\vert _{x^{\prime
}}~dx^{\prime}\right\vert \leq\delta_{\rho}g\left(  y\right)  \,,
\end{equation}
for every $\rho>0$ and some $\delta_{\rho}<\infty$. For the case $d=2,3$, it
is found that
\begin{align}
\epsilon^{2}N_{\pm}^{(2)}(\epsilon) &  =\frac{\pi}{4}\pm\epsilon
+\frac{\epsilon^{2}}{4}+\tilde{O}(\epsilon^{\tilde{w}_{2}}) \, , 
\label{Brownell2}
\\
\epsilon^{3}N_{\pm}^{(3)}(\epsilon) &  =\frac{\pi}{6}\pm\frac{3\pi}{8}%
\epsilon+\frac{3}{4}\epsilon^{2}\pm\frac{\epsilon^{3}}{8}+\tilde{O}%
(\epsilon^{\tilde{w}_{3}}) \, ,
\label{Brownell3}%
\end{align}
where $\tilde{w}_{2}>2$ and $\tilde{w}_{3}>3$ \cite{Baltes e Kn, Brownell}. As
will be seen below, the corrections~\eqref{Brownell2}-\eqref{Brownell3} are given by a
polynomial expansion whose coefficients are related to the geometric
\textquotedblleft shape\textquotedblright\ of the $\tilde{\Omega}_{d}$
boundary. Based on this geometric argument, it will be seen that we can always
decompose $\epsilon^{d}N_{+}^{(d)}(\epsilon)$ as the sum of two contributions:
one continuous component, that we will denote by $\mathcal{F}^{(d)}(\epsilon
)$, and other non-continuous.

Our goal in this section is to explore the relation between the functional
form of $\mathcal{F}^{(d)}(\epsilon)$ and the corrections derived from the
Brownell formalism. We will employ a \textquotedblleft bottom-up
approach\textquotedblright, considering in detail particular values for $d$,
and then extrapolating the results for general $d$.

The one-dimensional case ($d=1$) is trivial, however it is crucial to
construct the bottom-up approach. For this case, the Brownell corrections
coincide with the exact corrections. From Eq.~\eqref{eqfuncon}, the counting
function for Neumann boundary condition is given by
\begin{equation}
\epsilon N_{+}^{\left(  1\right)  }\left(  \epsilon\right)  =1-\epsilon
\eta\left(  \epsilon^{-1}\right)  \sim1\Rightarrow\mathcal{F}^{(1)}%
(\epsilon)=1\,. \label{EQN1+}%
\end{equation}
For the Dirichlet boundary condition the result is similar:
\begin{equation}
\epsilon N_{-}^{\left(  1\right)  }\left(  \epsilon\right)  =\epsilon\lbrack
N_{+}^{\left(  1\right)  }\left(  \epsilon\right)  -1]=1-\epsilon\lbrack
1+\eta\left(  \epsilon^{-1}\right)  ]\sim1\,.
\end{equation}

\begin{figure}
[ptb]
\begin{center}
\includegraphics[
height=2.6219in,
width=2.2751in
]%
{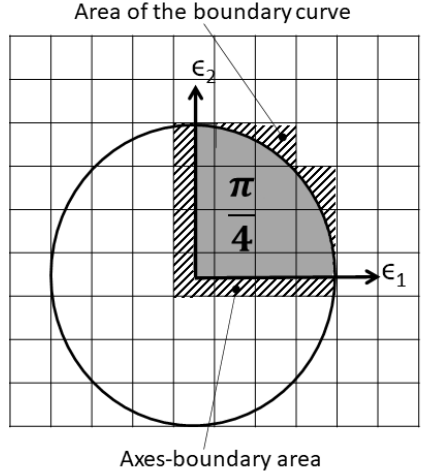}%
\caption{Two-dimensional coarse graining and lattice considered in the
counting process. The solid gray region contributes with an area
equals to $\pi/4$ (one fourth of the circle) and each axis contributes with an
area equals to $\epsilon/2$.}%
\label{fig1}%
\end{center}
\end{figure}

For the two-dimensional Neumann scenario, given the shape of $\tilde{\Omega
}_{2}$ boundary, we can decompose $\epsilon^{2}N_{+}^{(2)}$ as
\begin{equation}
\epsilon^{2}N_{+}^{(2)}(\epsilon)=\frac{\pi}{4}+\text{axes-boundary
area}+\text{area of the boundary curve}\,. \label{2-d}%
\end{equation}
This result is illustrated in Fig.~\ref{fig1}. More precisely, the
\textquotedblleft axes-boundary area\textquotedblright\ is the excess area
localized around the axes (which are not included in $\tilde{\Omega}_{2}$). As
seen in Fig.~\ref{fig1}, this quantity is
\begin{equation}
\text{axes-boundary area}=\epsilon+\frac{\epsilon^{2}}{4}\,.
\end{equation}
Notice that we are considering the contribution of each axis in the interval
$\left[  0,1\right]  $, implying that the contribution of the central square
is $\epsilon{}^{2}/4$. The contribution of the area of the boundary curve is
unknown.\footnote{This is related to the famous, and still open, Gauss circle
problem \cite{Emil}.} However we know that it is described by a non-continuous
function, otherwise $N_{+}^{(2)}(\epsilon)$ would be a continuous function and
we know that this is not true. Hence, it follows that $\mathcal{F}%
^{(2)}(\epsilon)$ in this case is
\begin{equation}
\mathcal{F}^{(2)}(\epsilon)=\frac{\pi}{4}+\epsilon+\frac{\epsilon^{2}}{4} \, ,
\label{eqgeo}%
\end{equation}
which agrees with the Brownell corrections given in Eq.~\eqref{Brownell2} for the Neumann case.

For the two-dimensional Dirichlet scenario, note that we can relate
$N_{-}^{(2)}$ and $N_{+}^{(2)}$ as%
\begin{equation}
N_{+}^{(2)}(\epsilon)=\underset{\text{points outside the axes}}{\underbrace
{N_{-}^{(2)}(\epsilon)}}+\underset{\text{points at the axes}}{\underbrace
{2N_{-}^{(1)}(\epsilon)+1}}\,.\label{eq NYD2}%
\end{equation}
Isolating $N_{-}^{(2)}$ and using results~\eqref{EQN1+} and \eqref{eqgeo}, we
obtain the Brownell corrections given in Eq.~\eqref{Brownell2} for Dirichlet.

\begin{figure}
[ptb]
\begin{center}
\includegraphics[
height=2.951in,
width=3.1514in
]%
{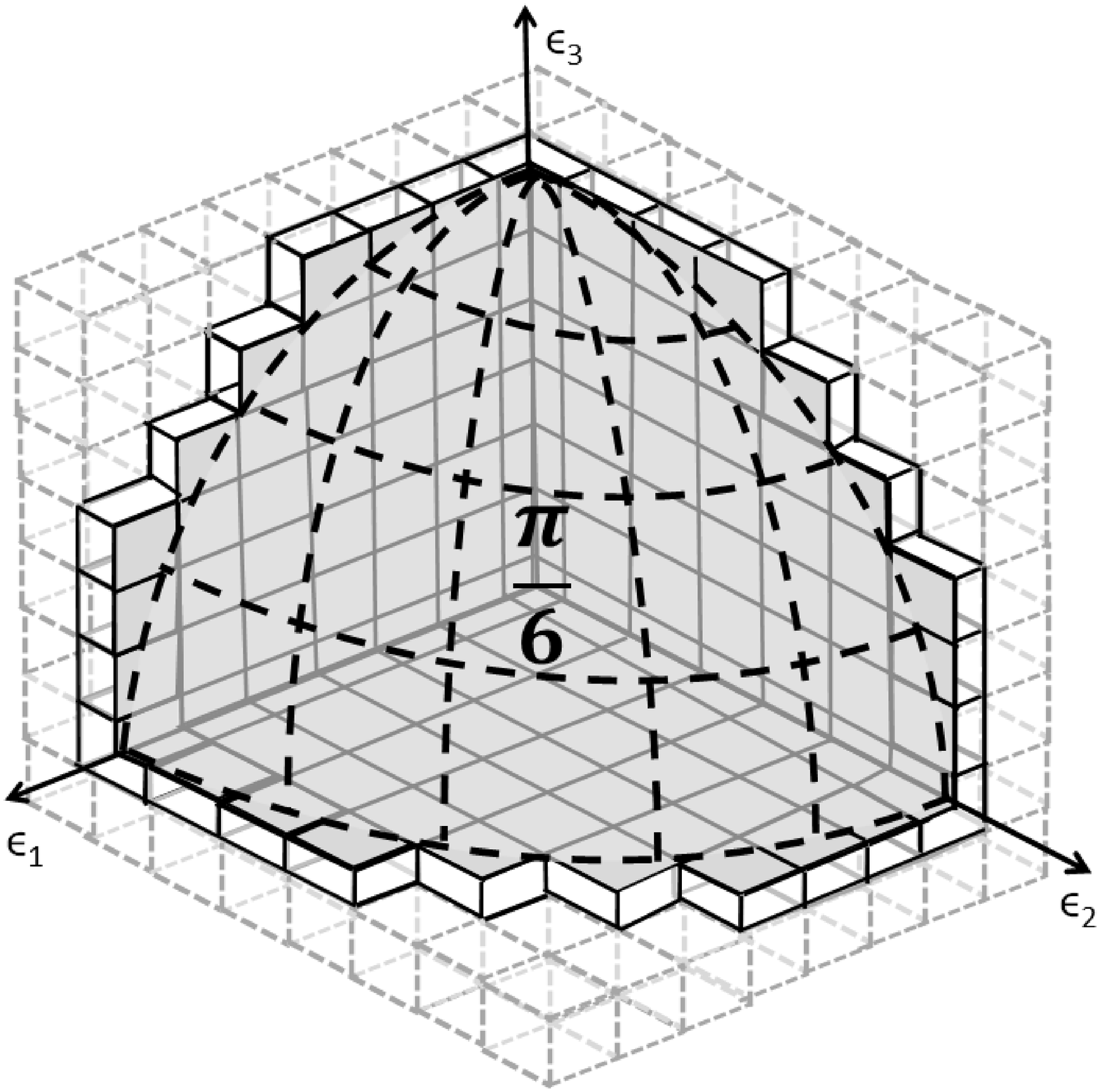}%
\caption{Three-dimensional coarse graining and axes-boundary area. The dotted dark
lines represent the $1/8$ of the sphere, that contributes with $\pi/6$. The
solids part of the cubes represents the axes-boundary area. The center cube
contributes with $\epsilon^{3}/8$. The cubes in each axis contributes with
$\epsilon^{2}/4 \,$. The cubes in the faces represents $3$ times $1/4$ of the
area of the circles, i.e., $3\pi/4$ times $\epsilon$. But only half this
value, i.e., $3\pi\epsilon/8$, contributes.}%
\label{fig2}%
\end{center}
\end{figure}

The analysis for the three-dimensional case can be conducted in an
analogous manner. To illustrate the development, let us consider
Fig.~\ref{fig2}, the generalization of two-dimensional lattice diagram (one
eighth of the three-dimensional sphere instead of one fourth of the
two-dimensional circle).

With the three-dimensional lattice with Neumann boundary conditions, we get
that $\mathcal{F}^{(3)}(\epsilon)$ for this case is
\begin{equation}
\mathcal{F}^{(3)}(\epsilon)=\frac{\pi}{6}+\frac{3\pi}{8}\epsilon+\frac{3}%
{4}\epsilon^{2}+\frac{\epsilon^{3}}{8}\,. \label{eqf3}%
\end{equation}
Again, the Brownell corrections for Neumann given in Eq.~\eqref{Brownell3}, are obtained.

For the three-dimensional case with Dirichlet boundary conditions, the
counting function can be decomposed as
\begin{equation}
N_{+}^{\left(  3\right)  }\left(  \epsilon\right)  =\underset{\text{points
outside the axes}}{\underbrace{N_{-}^{\left(  3\right)  }\left(
\epsilon\right)  }}+\underset{\text{points at the faces}}{\underbrace
{3N_{-}^{\left(  2\right)  }\left(  \epsilon\right)  }}+\underset{\text{points
at the axes}}{\underbrace{3N_{-}^{\left(  1\right)  }\left(  \epsilon\right)
+1}}\,.\label{nd-3d}%
\end{equation}
Using previous results, we obtain expression~\eqref{Brownell3}.

After studying the cases $d=2$ and $d=3$ in detail, we can generalize these results
to higher dimensions. Let us initially consider the $d$-dimensional scenario
with Neumann boundary conditions. As in Eq.~\eqref{2-d}, the counting function
$\epsilon^{d}N_{+}^{(d)}$ can be expressed as the sum of three contributions:
the term $|\tilde{\Omega}_{d}|$, the hypervolume associated to the axes and
the hypervolume of the boundary hypersurface. In appendix~\ref{eixos}, we show
that the hypervolume associated to the axes is given by
\begin{equation}
\text{Hypervolume of the axes}=\sum_{n=1}^{d}{\binom{d}{n}2^{-d}\omega
_{d-n}\epsilon^{n}}\,,
\end{equation}
where
\begin{equation}
{\binom{d}{n}=\frac{d!}{n!\left(  d-n\right)  !}~.}%
\end{equation}
Hence,
\begin{equation}
\mathcal{F}^{\left(  d\right)  }\left(  \epsilon\right)  =\sum_{n=0}%
^{d}{\binom{d}{n}2^{-d}\omega_{d-n}\epsilon^{n}~}\,. \label{eQ39}%
\end{equation}
The result obtained above generalizes the corrections~\eqref{Brownell2}-\eqref{Brownell3} to
an arbitrary dimension. By repeating the procedure realized in \cite{Brownell}
for $d=2,3$, and the using the Theorem (8.17) on this reference, we can find%
\begin{equation}
\epsilon^{d}N_{+}^{(d)}(\epsilon)=\mathcal{F}^{(d)}\left(  \epsilon\right)
+\tilde{O}(\epsilon^{\tilde{w}_{d}}) \, , \tilde{w}_{d}>d \, .
\end{equation}

The Dirichlet case can be obtained using a generalization of expressions~\eqref{eq NYD2} and \eqref{nd-3d}. This generalization is constructed in
Appendix~\ref{pascal}, and the final expression can be condensed into
\begin{equation}
\epsilon^{d}N_{\pm}^{\left(  d\right)  }\left(  \epsilon\right)  =\sum
_{n=0}^{d}\left[  \left(  -1\right)  ^{\frac{d-n}{2}}\right]  ^{1\mp1}%
{\binom{d}{n}2^{-d}\omega_{d-n}\epsilon^{n}}+\tilde{O}\left(  \epsilon
^{\tilde{w}_{d}}\right)  \, . \label{eqNDe2}%
\end{equation}
Expression~\eqref{eqNDe2} is the main result of this section. Note that this equation also maintains the \textquotedblleft alternating
symmetry\textquotedblright\ observed for the cases $d=2,3$, when we go from
Neumann to Dirichlet. If the coefficients do not follow this binomial pattern,
the alternating symmetry in the signs of coefficients is broken.

Rewriting expression~\eqref{eqNDe2} in terms of the variable $k$, we have the
results provided by Brownell's formalism for higher orders,
\begin{equation}
N_{\pm}^{\left(  d\right)  }\left(  k\right)  =\frac{1}{2^{d}}\sum_{n=0}%
^{d}\left[  \left(  -1\right)  ^{\frac{d-n}{2}}\right]  ^{1\mp1}{\binom{d}%
{n}\frac{\pi^{(n-d)/2}L^{d-n}}{\Gamma\left(  \frac{d-n}{2}+1\right)  }k^{d-n}%
}+\tilde{O}\left(  k^{d-\tilde{w}_{d}}\right)  \, , \label{nk}%
\end{equation}
where $\tilde{w}_{d}>d$.

\section{\label{comment}Mixed boundary conditions and degeneracies}

From the results involving Neumann and Dirichlet cases discussed, it is
possible to treat more complex boundary conditions, dubbed here as
\textquotedblleft mixed boundary conditions\textquotedblright. Also, in the
present section, we will consider the effects of degeneracies in the spectra.
Those setups model physical systems described by Helmholtz equation which will
be considered in this work. Further interesting examples (not directly
addressed here) can be found in \cite{Hacyan,Actor}.

In the present development, mixed boundary conditions are constructed imposing
Neumann and Dirichlet conditions in different axes. The solutions of Helmholtz
equation~\eqref{helmholtz d dimensoes} under these conditions can be written
as
\begin{equation}
\psi_{M}\left(  x_{1},\ldots,x_{d}\right)  =\phi^{\pm}\left(  x_{1}\right)
\cdots\phi^{\pm}\left(  x_{d}\right)  \,. \label{hibridcond}%
\end{equation}

In $d$ dimensions, the number of possible scenarios with mixed boundary
conditions is $2^{d}$, including the ones which are entirely Dirichlet or
Neumann. In any case, the counting function (referred generally as
$N_{M}^{(d)}$ for an arbitrary boundary condition) can be decomposed as a
linear combination in the form
\begin{equation}
N_{M}^{\left(  d\right)  }(\epsilon)=\sum_{n=0}^{d}A_{n}N_{-}^{\left(
n\right)  }\left(  \epsilon\right)  \,,\label{homomespect}%
\end{equation}
where the coefficients $\{A_{n}\}$ are related to the boundary conditions and
possible degeneracies. For example, result~\eqref{nm} for the Neumann case
without degeneracies is recovered with $A_{n}=\binom{d}{n}$ (see
Appendix~\ref{pascal} for details). This expression~\eqref{nm} can be
generalized with the consideration of vector fields associated to the
Helmholtz equation. This development will be necessary to the treatment of
degeneracies. In the vector-field problem, beside the $d$ modes $\{n_{1}%
,n_{2},\ldots,n_{d}\}$, there are also $d$ components for the fields,
$\{F_{n_{1}\cdots n_{d}}^{i}\,,\,i=1,\ldots,d\}$. However, contrary to the
scalar field, the internal degrees of freedom of the vector fields must be
taken into account. There is, the polarization of the vector field is an issue.

Possible polarizations can be transverse and longitudinal. Indeed, the vector
fields can be expressed as
\begin{equation}
F_{n_{1}\cdots n_{d}}^{i}=F^{0i}\prod_{k=1}^{\chi}\phi_{n_{k}}^{+}%
\prod_{j=\chi+1}^{d}\phi_{n_{j}}^{-} \, , \,\, i=1,\ldots,d\,, \label{1}%
\end{equation}
with constants $F^{0i}\in\mathbf{\mathbb{C}}$. Following the development which
leaded to Eq.~\eqref{nm}, non-trivial solutions can be separated into
$(\chi+1)$ classes: those in which only one $n_{i}$ is null, those in which
two $n_{i}$ are null, and so on, up to those which $\chi$ of the $n_{i}$ are
equal to zero. Let us label these classes as \textquotedblleft class
zero\textquotedblright, \textquotedblleft class one\textquotedblright, and so
on, respectively.

The solutions where $\chi$ of the possible $n_{k}$ are null (forming the class
$\chi$) have only one non-null component $F_{n_{1}\cdots n_{d}}^{i}$. These
solutions are polarized in the direction of this component. In the same way,
solutions with $(\chi-1)$ of the possible $n_{i}$ being null have two non-null
components, and consequently two possible polarizations. Hence, the class $n$
has a degenerescence equal to
\begin{equation}
\xi_{\chi-i}^{\left(  d,\chi\right)  }=i+1\,,\,\,\,i=0,\ldots,\chi
-1\,,\,\,\,\chi\leq d\,. \label{xmd}%
\end{equation}
For the class zero, where none of the $n_{i}$ is null, all the components
$F_{n_{1}\cdots n_{d}}^{i}$ can be non-null and we have $d$ degrees of
freedom, $\xi_{0}^{\left(  d,\chi\right)  }=d\,$. This counting of the degrees
of freedom considers only the effect of the boundary conditions. In addition,
one can also impose the orthogonality condition, which prevents the
longitudinal modes. This condition requires that the amplitude vector to be
perpendicular to the wave vector,
\begin{equation}
\sum_{n=1}^{d}k_{n}F^{0n}=0\,,
\end{equation}
which reduces by one the number of degrees of freedom of the zero class
configuration. Therefore, we can write
\begin{equation}
\xi_{0}^{\left(  d,\chi\right)  }=\xi_{0}^{\left(  d\right)  }=d-\tilde{\xi
}\,, \label{x2}%
\end{equation}
where $\tilde{\xi}=1$ for systems which admit only transverse perturbations
and $\tilde{\xi}=0$ for cases where longitudinal perturbations are considered.

As commented in Appendix~\ref{pascal}, each class $n$ of the $(\chi+1)$ vector
solutions behaves as the Dirichlet problem in $(d-n)$ dimensions. The class
with only one of the $n_{i}$ equal to zero (degenerescence equal to $\xi
_{1}^{\left(  d,\chi\right)  }$) can be divided in $d$ sub-classes ($d$ cases
where $n_{j}=0$). If two of the solutions have $n_{j}=0$ (degenerescence
$\xi_{2}^{\left(  d,\chi\right)  }$), the subclass can then be further
separated in $\binom{d}{2}$ sets. The process continues, until eventually new
sub-classes can not be produced. Given the counting process presented, the
total number of modes can be expressed as the composition of the classes and
sub-classes thus constructed,
\begin{equation}
N_{\chi}^{\left(  d\right)  }=\sum_{n=0}^{\chi}\left(  \#n\text{-class}%
\right)  \times\left(  n\text{-degenerescence}\right)  \times N_{-}^{\left(
d-n\right)  }\,,\label{2}%
\end{equation}
or
\begin{equation}
N_{\chi}^{\left(  d\right)  }\left(  \epsilon\right)  =\sum_{n=0}^{\chi}%
\binom{d}{n}\xi_{n}^{(d,\chi)}N_{-}^{\left(  d-n\right)  }\left(
\epsilon\right)  \,, 
\label{nk0}%
\end{equation}
with $N_{-}^{\left(  m\right)  }$ given by Eq.~\eqref{nk}, $\xi_{m}^{(d,\chi
)}$ in Eq.~\eqref{xmd} for $m>0$ and $\xi_{0}^{(d,\chi)}$ in Eq.~\eqref{x2}.
We emphasize that expression~\eqref{nk0} is one of the main results of the
present work. For $\chi=d$ and no polarization ($\xi_{i}^{\left(  d\right)
}=1\,,\ \forall i=0,\ldots,d$), Eq.~\eqref{nm} in Appendix~\ref{pascal} is recovered.

In the following sections we will apply the formalism developed in concrete
scenarios. Namely, we will discuss the thermodynamics of the electromagnetic
field in a hypercubic cavity and acoustic perturbations in a generalized version of Debye model.

\section{\label{thermodynamics} Thermodynamic and quasithermodynamic limits}

From the Weyl law and its extensions, we turn to physics applications. The
goal of this section is to connect the derived expansions for the counting
functions with thermodynamic analyses. In this context, corrections of the
Weyl law will correspond to the transition from the strict thermodynamic limit
to the quasithermodynamic approach.

Let us assume a semi-classical treatment, with the physical system of interest being described by solutions
of the Helmholtz equation~\eqref{helmholtz d dimensoes}. In the treatment, we consider a $d$-dimensional cubic box with side length $L$ populated by a thermal gas. The gas is composed by effective massless particles (actually modes associated to electromagnetic or acoustic perturbations) with a well-defined energy and subjected
to the Bose-Einstein statistics. The system is supposed to be in thermal
equilibrium with constant temperature $T$. The number of modes is not
conserved, and hence the chemical potential $\mu$ is null. Since the
temperature and chemical potential are fixed, the grand canonical ensemble is assumed.

A macroscopic treatment is established if a thermodynamic limit can be
achieved. Since that we are employing a grand canonical ensemble, the strict
thermodynamic limit is defined as \cite{Kuzemsky}
\begin{equation}
|\Omega_{d}|\rightarrow\infty\,\,\text{with}\,\,T\,\,\text{fixed}%
\,\,\text{and}\,\,\mu=0\,\,, \label{thermodynamic_limit}%
\end{equation}
where $|\Omega_{d}|$ is the hypervolume of the domain $\Omega_{d}$. As seen
from Eq.~\eqref{volume do dominio dD}, this condition implies that
\begin{equation}
LT\rightarrow\infty\,\,\text{with}\,\,T\neq0\,\,\text{fixed}\,\,\text{and}%
\,\,\mu=0\,. \label{thermodynamic_limit_2}%
\end{equation}

It is also important to consider not only the thermodynamic limit, but also
how this limit is approached. Hence, we establish the quasithermodynamic
limit~\cite{Maslov1} as
\begin{equation}
|\Omega_{d}|\,\,\text{large but finite, with}\,\,T\,\,\text{fixed}%
\,\,\text{and}\,\,\mu=0\,\,.
\end{equation}
The quasithermodynamic is particularly relevant in the present work, where
finite cavities or solids are considered.

Given that the system of interest is compatible with thermodynamic and
quasithermodynamic limits, the associated partition function can be written
as
\begin{equation}
\ln Z=-\int_{0}^{\infty}\,D(\omega)\,\ln\left[  1-\exp\left(  -\frac
{\hbar\omega}{k_{B} T}\right)  \right]  \,\mathrm{d}\omega\,,
\label{Z-general}%
\end{equation}
where $D(\omega)$ is the spectral density function.
Expression~\eqref{Z-general} can be seen as the Thomas-Fermi approximation for
the exact grand canonical partition function. From $Z$, all the associated
thermodynamic and quasithermodynamic quantities can be readily calculated. For
example, the internal energy is given by
\begin{equation}
U = k_{B} T^{2}\frac{\partial}{\partial T}\,\left(  \ln Z\right)  \,.
\label{def_U}%
\end{equation}

In a practical implementation of the quasithermodynamic limit, combined with
the assumption that the temperature of the system is fixed, the box length $L
$ should be large enough so that \cite{Elias:2018yct}
\begin{equation}
\frac{k_{B}}{\hbar c} \, LT \gg1 \,, \,\, \text{with $LT$ kept finite.}
\label{quasithermodynamic_limit}%
\end{equation}
As reference, $k_{B} / (\hbar c) \approx436.7 \, \text{K}^{-1} \,
\text{m}^{-1}$ in SI units. The relation between the spectral density $D$ in
the partition function~\eqref{Z-general} and the counting functions studied in
the previous sections (collectively denoted by $N^{(d)}$) is given by
\begin{equation}
D(\omega)\,\mathrm{d}\omega=\frac{\mathrm{d}N^{(d)}(\omega)}{\mathrm{d}\omega
}\,\mathrm{d}\omega\,.
\end{equation}
It follows that the Weyl law \eqref{weyl-law} is connected with the strict
thermodynamic limit~\eqref{thermodynamic_limit_2} when only the term with
highest power of $LT$ is considered in the asymptotic expansion of the
counting function. For the link between extensions of Weyl law and the
quasithermodynamic limit~\eqref{quasithermodynamic_limit}, subdominant
corrections on powers of $LT$ are also considered in this expansion.
It should be noted that, although result~\eqref{nk0} for the counting function is valid for any values of $L$, it is only when condition~\eqref{quasithermodynamic_limit} is satisfied that the integral form~\eqref{Z-general} of the partition function can be employed.

\section{\label{photons} Quasithermodynamics of the electromagnetic field}

A first application of the developed formalism involves quasithermodynamics of
the electromagnetic field in a (hyper)cubic cavity. Let us consider a cubical
cavity in $d$ dimensions with side length $L$. Its faces are supposed to be
perfect conductors, surrounding a vacuum region. The electric field inside the
cavity satisfies a wave equation characterized by a velocity equals to $c$.
Also, the conductivity of the walls guaranties that the tangential components
of the electric field at the walls are null.

The components of the electric field inside the cavity are given by
\cite{Liverpool}
\begin{equation}
E_{n_{1}\cdots n_{d}}^{i}\left(  x_{1},\ldots,x_{d}\right)  =E^{0i}\times
\phi_{n_{1}}^{-}\left(  x_{1}\right)  \cdots\phi_{n_{i}}^{+}\left(
x_{i}\right)  \cdots\phi_{n_{d}}^{-}\left(  x_{d}\right)  \,,\label{Ecamp}%
\end{equation}
where each component satisfies a mixed boundary
condition~\eqref{1} with $\chi=1$. Furthermore, as only transverse modes are
present, we have $\tilde{\xi}=1$ in Eq.~\eqref{x2}. Hence, using
result~\eqref{nk0}, we find that the number of independent modes
$N_{\text{em}}^{\left(  d\right)  }=N_{1}^{\left(  d\right)  }$ is given by
\begin{equation}
N_{\text{em}}^{\left(  d\right)  }\left(  \omega\right)  =\frac{1}{2^{d}%
}\left[  \frac{\left(  d-1\right)  }{c^{d}\pi^{d/2}}\frac{\left(  \omega
L\right)  ^{d}}{\Gamma\left(  \frac{d}{2}+1\right)  }+\frac{d\left(
3-d\right)  }{c^{d-1}\pi^{\left(  d-1\right)  /2}}\frac{\left(  \omega
L\right)  ^{d-1}}{\Gamma\left(  \frac{d-1}{2}+1\right)  }\right]
\,.\label{n modos EM correcao 1}%
\end{equation}
In Eq.~\eqref{n modos EM correcao 1}, it was assumed the vacuum dispersion
relation $k=\omega/c$ for the electromagnetic field.

For the three-dimensional case we have the well-known quadratic term
cancellation in $\omega$ \cite{Baltes Baltes, Liu}. In this case, we consider
the next correction term in Eq.~\eqref{nk0}, furnishing
\begin{equation}
N_{\text{em}}^{\left(  3\right)  }=\frac{L^{3}}{3\pi^{2}c^{3}}\omega^{3}%
-\frac{3L}{2\pi c}\omega\,. \label{N-elect-d3}%
\end{equation}
Result~\eqref{N-elect-d3} agrees with \cite{Baltes Baltes, Baltes e Kn, Baltes
S,Liu}. However, Eq.~\eqref{n modos EM correcao 1} shows that this
cancellation of the \textquotedblleft area\textquotedblright\ term for the
electromagnetic field only occurs in the three-dimensional case. This implies
a drastic difference in $d=3$ scenario when comparing with other dimensionalities.

The spectral density $D_{\text{em}}^{\left(  d\right)  }$ can be calculated
deriving expression~\eqref{n modos EM correcao 1} with respect to $\omega$,
\begin{equation}
D_{\text{em}}^{\left(  d\right)  }=\frac{\mathrm{d}N_{\text{em}}^{\left(
d\right)  }}{\mathrm{d}\omega}=\frac{1}{2^{d}}\left[  \frac{d\left(
d-1\right)  L^{d}\omega^{d-1}}{\pi^{d/2}\Gamma\left(  \frac{d}{2}+1\right)
c^{d}}+\frac{d\left(  d-1\right)  \left(  3-d\right)  L^{d-1}\omega^{d-2}}%
{\pi^{\left(  d-1\right)  /2}\Gamma\left(  \frac{d-1}{2}+1\right)  c^{d-1}%
}\right]  \,. \label{D-em}%
\end{equation}
Hence, the internal energy of the system can be written as
\begin{equation}
U^{\left(  d\right)  }=\int_{0}^{\infty}D_{\text{em}}^{\left(  d\right)
}\left(  \omega\right)  \frac{\hbar\omega}{\exp\left(  \frac{\hbar\omega
}{k_{B}T}\right)  -1}~\mathrm{d}\omega=k_{B}\left[  \left(  d-1\right)
\theta_{d}L^{d}T^{d+1}+\left(  3-d\right)  \frac{d}{2}\theta_{d-1}L^{d-1}%
T^{d}\right]  \,, \label{new stefan d}%
\end{equation}
where $\theta_{m}$ is defined as
\begin{equation}
\theta_{m}\equiv\frac{\zeta\left(  m+1\right)  \Gamma\left(  m+1\right)
m}{2^{m}\pi^{m/2}\Gamma\left(  \frac{m}{2}+1\right)  }\frac{k_{B}^{m}}%
{\hbar^{m}c^{m}}\,. \label{theta-n}%
\end{equation}
The term $\zeta\left(  z\right)  $ in Eq.~\eqref{theta-n} denotes the Riemann
zeta function. Expression~\eqref{new stefan d} can be seen as an improved
version of the Stefan-Boltzmann law for the electromagnetic field in a
hypercubic cavity, in a quasithermodynamic treatment.

In the strict thermodynamic limit, that is, when the correction is not
considered, we recover the Stefan-Boltzmann law in $d$ dimensions. In the
strict thermodynamic regime, the (improved) result~\eqref{new stefan d} can be
compared with \cite{Landsberg, Alnes}. However, in those references the
authors impose that $\xi_{\text{em}}^{(d)}=\xi_{0}^{\left(  d\right)  }=2$ for
the polarization of the $d$-dimensional electric field (as done for the
three-dimensional scenario). The observation that $\xi_{\text{em}}^{(d)}=d-1$
is the correct factor was made in \cite{Comment}.

It is important to stress that, while the enforcement that $\xi_{\text{em}%
}^{(d)}=2$ for any dimension corresponds to a simple factor for the main term
of several thermodynamic quantities, this choice has a very drastic effect for
the correction terms in the quasithermodynamic analysis. For instance, when
considering the quasithermodynamics, the general enforcement of $\xi
_{\text{em}}^{(d)}=2$ would imply the cancellation of the first correction for
any value of $d$ (not appropriate), and not for $d=3$, as we have obtained.

The quasithermodynamic Stefan-Boltzmann law~\eqref{new stefan d} can also be
rewritten as
\begin{equation}
\frac{U^{\left(  d\right)  }}{L^{d}}=\int_{0}^{\infty}B^{\left(  d\right)
}\left(  \omega,T,L\right)  ~\mathrm{d}\omega\,,
\end{equation}
implying that
\begin{equation}
B^{\left(  d\right)  }\left(  \omega,T,L\right)  =\left[  \frac{\omega}%
{\Gamma\left(  \frac{d}{2}+1\right)  }-\frac{c\left(  d-3\right)  \sqrt{\pi}%
}{\Gamma\left(  \frac{d-1}{2}+1\right)  }\frac{1}{L}\right]  \frac{d\left(
d-1\right)  \hbar\omega^{d-1}}{2^{d}\pi^{d/2}c^{d}\left[  \exp\left(
\frac{\hbar\omega}{k_{B} T} \right)  -1\right]  }\,. \label{New Planck}%
\end{equation}
Expression~\eqref{New Planck} is a quasithermodynamic version of the Planck
formula. This result describes the effects of the boundedness of the cavity on
the spectral density of the electromagnetic field.

For the two-dimensional case, a more detailed discussion is in order. Our
results for $d=2$ can be compared with the analysis presented in \cite{stefan
2d}. An interesting remark suggested in \cite{stefan 2d} is that size effects
prevent arbitrarily low frequencies in the system. With this observation,
particularized here for a square system of area $L^{2}$, the authors propose
an internal energy $U$ with the form
\begin{equation}
U=2\int_{\omega_{\min}}^{\infty}D\left(  \omega\right)  \frac{\hbar\omega
}{\exp\left(  \frac{\hbar\omega}{k_{B} T} \right)  -1}~\mathrm{d}\omega\, ,
\,\, D\left(  \omega\right)  =\frac{L^{2}\omega}{2\pi c^{2}} \, , \,\,
\omega_{\min}=\frac{\sqrt{2} \pi c}{L}\,. \label{E min}%
\end{equation}
The integral in Eq.~\eqref{E min} can be solved by making the change
$x=t+x_{\min}$, where
\begin{equation}
x \equiv\frac{\hbar\omega}{k_{B} T} \, , \,\, x_{\min} \equiv\frac{\hbar
\omega_{\min}}{k_{B} T} =\frac{\sqrt{2}\pi\hbar c}{k_{B}TL}\,. \label{x min}%
\end{equation}
One obtains%
\begin{align}
&  U\left(  x_{\min}\right)  = \frac{k_{B}^{3}}{\pi\hbar^{2}c^{2}} \, L^{2}
T^{3} S\left(  x_{\min}\right)  \, ,\\
&  S\left(  x_{\min}\right)  =2\mathrm{Li}_{3}\left(  e^{-x_{\min}}\right)
+2x_{\min}\mathrm{Li}_{2}\left(  e^{-x_{\min}}\right)  +x_{\min}%
^{2}\mathrm{Li}_{1}\left(  e^{-x_{\min}}\right)  \,, \label{kimU}%
\end{align}
where the polylogarithm function $\mathrm{Li}_{s}\left(  z\right)  $ is
defined as
\begin{equation}
\mathrm{Li}_{s+1}\left(  z\right)  =\frac{1}{\Gamma\left(  s+1\right)  }%
\int_{0}^{\infty}\frac{t^{s}}{\exp\left(  t\right)  /z-1}~\mathrm{d}t\,.
\end{equation}
Following the development in \cite{stefan 2d}, the term $S\left(  x_{\min
}\right)  $ is expanded around $x_{\min}=0$,%
\begin{equation}
S(x_{\min})=2\zeta(3)-\frac{x_{\min}^{2}}{2}+\frac{x_{\min}^{3}}{6}%
-\frac{x_{\min}^{4}}{48}-\frac{x_{\min}^{6}}{4320}+O\left(  x_{\min}%
^{7}\right)  \,. \label{kimS}%
\end{equation}
Observe that the expansion close to $x_{\min}=0$ is equivalent to consider
$\left[  k_{B}/(\hbar c) \right]  \, LT \gg1$, as one sees from the definition
of $x_{\min}$ in Eq.~\eqref{x min}. Finally, using the expansion~\eqref{kimS},
and keeping only the first-order term $x_{\min}^{2}$, the approach from
\cite{stefan 2d} furnishes\footnote{It should be remarked that
Eq.~\eqref{U do cara} is not actually derived in \cite{stefan 2d}.}
\begin{equation}
U=\frac{2\zeta\left(  3\right)  }{\pi}\frac{k_{B}^{3}}{\hbar^{2}c^{2}}%
L^{2}T^{3}-\pi k_{B}T\,. \label{U do cara}%
\end{equation}

We note that the first term in Eq.~\eqref{U do cara} differs from the
result~\eqref{new stefan d} presented in this work by a factor of $2$. This
discrepancy is just a consequence of the authors in \cite{stefan 2d} using the
developments of \cite{Landsberg} which, as previously commented, assume that
$\xi_{\text{em}}^{(d)}=2$ for any value of $d$. However, as a more important
remark, it is possible to see from Eq.~\eqref{E min} that, while considering
the size effects on the minimum frequency, these effects on the density of the
modes $D\left(  \omega\right)  $ are not taken into account. In other words,
the integrand of $U$ in Eq.~\eqref{E min} is correct only in the thermodynamic limit.

We propose an improvement of the results in \cite{stefan 2d} by considering
size effects for both the minimal frequency and the density of the modes.
Following the procedure that leads to Eq.~\eqref{kimU}, but using
$D_{\text{em}}^{\left(  2\right)  }$ of Eq.~\eqref{D-em} instead of
$D(\omega)$ in the integrand of the internal energy $U$ in
expression~\eqref{E min}, we get
\begin{equation}
U_{\text{em}}^{\left(  2\right)  }=\pi k_{B}T\left[  x_{\min}^{-2}S\left(
x_{\min}\right)  +\frac{\sqrt{2}}{\pi}x_{\min}^{-1}\tilde{S}\left(  x_{\min
}\right)  \right]  \, , \label{nossoU}%
\end{equation}
with $S(x_{\min})$ presented in Eq.~\eqref{kimU} and
\begin{equation}
\tilde{S}\left(  x_{\min}\right)  =\mathrm{Li}_{2}\left(  e^{-x_{\min}%
}\right)  +x_{\min}\mathrm{Li}_{1}\left(  e^{-x_{\min}}\right)  \,.
\label{S-til}%
\end{equation}
The expansion of $\tilde{S}(x_{\min})$ around $x_{\min}=0$ gives
\begin{equation}
\tilde{S}\left(  x_{\min}\right)  =\frac{\pi^{2}}{6}-x_{\min}+\frac{x_{\min
}^{2}}{4}-\frac{x_{\min}^{3}}{36}+\frac{x_{\min}^{5}}{3600}+O\left(  x_{\min
}^{6}\right)  \,. \label{nossoS}%
\end{equation}
Substituting the results~\eqref{kimS} and \eqref{nossoS} in Eq.~(\ref{nossoU}%
), and keeping only terms of order up to $x_{\min}^{2}$, we obtain
\begin{equation}
U_{\text{em}}^{\left(  2\right)  }=\frac{\zeta\left(  3\right)  }{\pi}%
\frac{k_{B}^{3}}{\hbar^{2}c^{2}}L^{2}T^{3}+\frac{\pi k_{B}^{2}}{6\hbar
c}LT^{2}-\left(  \frac{1}{2}+\frac{\sqrt{2}}{\pi}\right)  \pi k_{B}T\,.
\label{Uem}%
\end{equation}
In addition to the already mentioned factor of $2$, and a correction of
$\sqrt{2}/\pi$ in the last term in Eq.~\eqref{U do cara}, we highlight the
appearance of an \textquotedblleft area\textquotedblright\ term proportional
to $T^{2}$. This new term is the leading correction on the quasithermodynamics limit.

It is important to notice that the second term in Eq.~\eqref{Uem}, that is,
the leading correction, is precisely the last term presented in
Eq.~\eqref{new stefan d} for $d=2$. This is a consequence of the fact that $S$
in Eq.~\eqref{kimS} does not have a linear term (i.e., a term proportional to
$x_{\min}$). The existence of such linear term would change the second term in Eq.~\eqref{new stefan d}.

The above procedure for the two-dimensional scenario can be generalized to
arbitrary dimensions. In this case, to consider a minimal energy implies that%
\begin{equation}
U_{\text{em}}^{\left(  d\right)  }=d\left(  d-1\right)  k_{B}T\left[  \left(
x_{\min}^{\left(  d\right)  }\right)  ^{-d}C_{d}S_{d}\left(  x_{\min}^{\left(
d\right)  }\right)  +\left(  x_{\min}^{\left(  d\right)  }\right)  ^{-1}%
\tilde{C}_{d}\tilde{S}\left(  x_{\min}^{\left(  d\right)  }\right)  \right]
\,, \label{kimud}%
\end{equation}
where
\begin{equation}
\begin{aligned} & x_{\min}^{\left( d\right) }=\frac{\hbar c\pi}{k_{B}LT}\sqrt{\frac{3d}{2}-1} \, , \,\, C_{d}=\left( \frac{\pi}{8}\right) ^{\frac{d}{2}}\frac{\left( 3d-2\right) ^{\frac{d}{2}}}{\Gamma\left( \frac{d}{2}+1\right) }\, , \,\, \tilde{C}_{d}=\frac{\left( -1\right) ^{d}}{{{2^{d-1}}}\pi}C_{1} \, , \\ & S_{d}\left( x_{\min}^{\left( d\right) }\right) =\sum_{k=0}^{d}\left( \begin{array} [c]{c}d\\ k \end{array} \right) \Gamma\left( k+1\right) \left( x_{\min}^{\left( d\right) }\right) ^{d-k}\mathrm{Li}_{k+1}\left( e^{-x_{\min}^{\left( d\right) }}\right) \,, \end{aligned}
\end{equation}
and $\tilde{S}$ is given by Eq.~\eqref{S-til} for any value of $d$. Similarly
to the two-dimensional case, it is straightforward to see that the limit of
$S_{d}$ around $x_{\min}^{\left(  d\right)  }=0$ does not have a linear term.

It should be noticed that the power of $x_{\min}^{\left(  d\right)  }$ in the
second term in the brackets of Eq.~\eqref{kimud} does not depend on dimension
and is equal to $-1$. Thus, the higher contribution of this term is
proportional to $LT^{2}$, in the same way as in $d=2$. Therefore, the leading
terms in $U_{\text{em}}^{\left(  d\right)  }$ of Eq.~\eqref{kimud} are
precisely those presented in expression~\eqref{new stefan d}. We conclude
that, for $d\neq3$, the existence of a minimal frequency would not affect the
quasithermodynamics behavior, exactly as in the thermodynamic limit, and the
expression~\eqref{new stefan d} is correct even if the minimal frequency is
considered.\footnote{For the three-dimensional scenario, the absence of the
area term could imply higher-order corrections. In
this case, the minimum frequency would affect the first correction term, which
is proportional to $LT^{2} \,$.}

\section{\label{phonons} Quasithermodynamics of acoustic perturbations}

\subsection{General considerations}

We turn to the quasithermodynamics of acoustic perturbations, focusing on the $d$-dimensional
version of the Debye model. Let us consider a harmonic solid, there is, an
isotropic, elastic and continuous body. The solid is assumed to be a hypercube
of dimension $d$ and length size $L$. In this hypercube, a number of $\chi$
opposed faces are free, and hence the oscillations in these directions respect
Neumann conditions. The remaining $d-\chi$ opposing faces are fixed,
respecting Dirichlet conditions. The propagation of vibrations in the solid is
associated to acoustic waves.

In this setup, the components of the displacement field $u^{i}(x)$ of the
particles that form the solid (atoms, ions, molecules, etc.) will be solutions
of Helmholtz equation with the form~\eqref{1}. Specifically, $u^{i}(x)$ are
given by
\begin{equation}
u_{n_{1}\cdots n_{d}}^{i}\left(  x_{1},\ldots,x_{d}\right)  =u^{0i}%
\times\underset{\text{free faces}}{\underbrace{\phi_{n_{1}}^{+}\left(
x_{1}\right)  \cdots\phi_{n_{\chi}}^{+}\left(  x_{\chi}\right)  }}%
\underset{\text{fixed faces}}{\underbrace{\phi_{n_{\chi+1}}^{-}\left(
x_{\chi+1}\right)  \cdots\phi_{n_{d}}^{-}\left(  x_{d}\right)  }}\,.
\label{uFON}%
\end{equation}

Contrary to the description of the electromagnetic field in a cavity, acoustic perturbations
are free to oscillate in the longitudinal direction. Therefore, a distinct
quasithermodynamic behavior, comparing to the thermodynamic of electromagnetic
perturbations, should be expected in the present setup. In particular, for
$d=3$, we should expect an important role of the area term, the first term of
quasithermodynamic correction.\footnote{Effects associated to the area term in
a three-dimensional setup with low temperature are considered in
\cite{Dupuis,montroll}.}

Let us apply the results derived in the present work to generalize the
well-known expressions in the usual three-dimensional Debye model and
investigate the quasithermodynamic regime. From Eq.~\eqref{nk0}, with
$\tilde{\xi}=0$ in~(\ref{x2}), we obtain
\begin{equation}
N_{s}^{\left(  d\right)  }\left(  \omega\right)  =\frac{d}{\pi^{d/2}2^{d}%
}\left[  \frac{1}{\Gamma\left(  \frac{d}{2}+1\right)  }\left(  \frac{\omega
L}{c_{s0}^{\left(  d\right)  }}\right)  ^{d}+\frac{\left(  2\chi-d\right)
\pi^{1/2}}{\Gamma\left(  \frac{d-1}{2}+1\right)  }\left(  \frac{\omega
L}{c_{s\chi}^{\left(  d\right)  }}\right)  ^{d-1}\right]  \,.
\label{N-phonons}%
\end{equation}
The quantities $c_{s0}^{\left(  d\right)  }$ and $c_{s\chi}^{\left(  d\right)
}$ represent an effective bulk sound velocities and will be discussed in the
next section.

Expression~\eqref{N-phonons} can be interpreted as the number of modes
associated to acoustic waves whose frequencies are lower than $\omega$. Unlike the
electromagnetic case, in the treatment of Debye model the area term is absent
only for $d$ even and when there is a precise balance between Neumann and
Dirichlet boundary conditions: half the faces of the solid are free and half
are fixed. In particular, for the three-dimensional scenario, there is always
the influence of the area term.

\subsection{Influence of the area term}

Let us focus on the effects of the area term, considering the extreme
scenarios, where all the walls are free ($\chi=d$), or all the walls are fixed
($\chi=0$). In these cases,
\begin{equation}
N_{s\pm}^{\left(  d\right)  }\left(  \omega\right)  =\frac{1}{\pi^{d/2}}%
\frac{d}{2^{d}}\left[  \frac{1}{\Gamma\left(  \frac{d}{2}+1\right)  }\left(
\frac{\omega L}{c_{s0}^{\left(  d\right)  }}\right)  ^{d}\pm\frac{d\pi^{1/2}%
}{\Gamma\left(  \frac{d-1}{2}+1\right)  }\left(  \frac{\omega L}{c_{s\pm
}^{\left(  d\right)  }}\right)  ^{d-1}\right]  \,. \label{Nsom}%
\end{equation}
In Eq.~\eqref{Nsom}, the plus and minus signs are associated to the
\textquotedblleft all free\textquotedblright\ and \textquotedblleft all
fixed\textquotedblright\ types of solid, respectively. Also, $c_{s0}^{\left(
d\right)  }$ can be interpreted as an effective velocity given by the linear
superposition of velocities $c_{l}$ and $c_{t}$, of the longitudinal mode and
of the $(d-1)$ transverse modes,
\begin{equation}
\left(  c_{s0}^{\left(  d\right)  }\right)  ^{d}=dc_{t}^{d} \left(
d-1+\frac{c_{t}^{d}}{c_{l}^{d}}\right)  ^{-1}\,.
\end{equation}
It is important to notice that this linear superposition can only be applied
to the main term \cite{Baltes e Kn}. Indeed, the reflection of the modes in
the walls of the solid produce a mixture of the modes. Hence, a purely
transverse (or longitudinal) perturbation can be reflected as a superposition
of transverse and longitudinal waves. The phenomenon generates an effective
velocity $c_{s\pm}^{\left(  d\right)  }$ which is different from
$c_{s0}^{\left(  d\right)  }$.

For the three-dimensional case, one approach to study the wave reflection in a
specific wall is to consider a slab instead of a cube, and hence to ignore
border effects in this wall. That is the approach followed in \cite{Dupuis},
where appropriate boundary conditions (Neumann or Dirichlet) are imposed on
the faces parallel to the plane of the slab (the planes of reflection).
However, periodic boundary conditions are enforced on the other faces (i.e.,
on the slab thickness direction), in an approach which captures border
effects. Note that the number of faces with periodic boundary condition is
equal to the dimension of the incident plane (two-dimensional in the
three-dimensional case). Periodic conditions at the borders are justified if
the borders are far enough from the region of interest. Within this simplified
model, it was determined in \cite{Dupuis} that
\begin{align}
&  \left(  c_{s+}^{\left(  3\right)  }\right)  ^{2}=3\left[  \frac{2\left(
c_{t}^{2}\right)  ^{2}-3c_{t}^{2}c_{l}^{2}+3\left(  c_{l}^{2} \right)  ^{2}%
}{c_{t}^{2}c_{l}^{2}\left(  c_{l}^{2}-c_{t}^{2}\right)  }\right]  ^{-1} \,
,\label{ud3-2}\\
&  \left(  c_{s-}^{\left(  3\right)  }\right)  ^{2}=3\left[  \frac{2}%
{c_{t}^{2}}+\frac{1}{c_{l}^{2}}+\frac{\left(  c_{l}^{2}-c_{t}^{2}\right)
^{2}}{c_{l}^{2}c_{t}^{2}\left(  c_{l}^{2}+c_{t}^{2}\right)  }\right]  ^{-1}\,.
\label{ud3}%
\end{align}
Following the development presented in \cite{Dupuis}, it is observed that the
presence of the power $2$ in $c_{l,t}$ is a consequence of the number of
periodic boundary conditions considered. That, in turn, is a result of the
fact that the plates have two dimensions.

In $d$ dimensions, we can consider the reflection of the wave by any one of
the $2d$ faces. In this case, a slab is defined as the set of all points lying
between two $(d-1)$-dimensional hyperplanes\footnote{See section 3.4 of
\cite{dg}.} in $\mathbb{R}^{d}$. In this way, the system is approximated by
two infinite plates. Appropriate boundary conditions (Neumann or Dirichlet)
are imposed on these plates, with periodic boundary conditions enforced on the
other $(d-1)$ directions. In this way, considering the relation with the
dimensionality of the system and the power series on $c_{t}$ and $c_{l}$, the
proposed generalization of the three-dimensional results in
Eqs.~\eqref{ud3-2}-\eqref{ud3} to the more general $d$-dimensional scenario is%
\begin{equation}
\left(  c_{s\pm}^{\left(  d\right)  }\right)  ^{d-1}=dc_{t}^{d-1}\left[
d-1+\frac{\left(  c_{l}^{d-1}\right)  ^{2}-c_{l}^{d-1}c_{t}^{d-1}+2\left(
c_{t}^{d-1}\right)  ^{2}}{c_{l}^{d-1}\left(  c_{l}^{d-1}\mp c_{t}%
^{d-1}\right)  }\right]  ^{-1}\,. \label{du}%
\end{equation}

\subsection{Debye frequency}

One important parameter of the Debye model is the so-called Debye frequency.
This quantity refers to the cutoff angular frequency of the waves propagating
in the solid, resulting from the existence of a minimum distance between the
particles that form the solid lattice. From Eq.~\eqref{Nsom}, it is possible
to determine the Debye frequency $\omega_{D\pm}^{\left(  d\right)  }$,
analyzing the number of degrees of freedom of the system. That is,
\begin{equation}
nd+n_{\partial}l\approx nd=N_{s\pm}^{\left(  d\right)  }\left(  \omega_{D\pm
}^{\left(  d\right)  }\right)  \,, \label{deb}%
\end{equation}
where $n$ is the number of particles inside the cavity (bulk) and
$n_{\partial}$ is the number of particles in the borders (edge). Considering
the edge, the particles have $l$ degrees of freedom. In
expression~\eqref{deb}, we assumed that $n\gg n_{\partial}$ since we are
performing a macroscopic (quasithermodynamic) treatment, and hence we can
consider $n$ as an approximation to the total number of particles of the
system. In fact, we are interested in the border effects on the modes
propagating on the bulk, not considering the superficial modes (Rayleigh
modes). The contribution of the superficial modes can be disregarded in the
quasithermodynamic regime.

One relevant issue is the influence of the borders in Debye frequency. For the
analysis of this point, we rewrite expression~\eqref{deb} in the form
\begin{equation}
\left(  \omega_{D\pm}^{\left(  d\right)  }\right)  ^{d}+B_{d\pm}\left(
\omega_{D\pm}^{\left(  d\right)  }\right)  ^{d-1}-\left(  \omega_{D0}^{\left(
d\right)  }\right)  ^{d}=0\,, \label{polinomial equation}%
\end{equation}
with
\begin{align}
&  B_{d\pm}\equiv\pm\frac{d\pi^{1/2}\Gamma\left(  \frac{d}{2}+1\right)
}{\Gamma\left(  \frac{d-1}{2}+1\right)  }\frac{\left(  c_{s0}^{\left(
d\right)  }\right)  ^{d}}{\left(  c_{s\pm}^{\left(  d\right)  }\right)
^{d-1}L}\,,\label{coeficiente Bd}\\
&  \omega_{D0}^{\left(  d\right)  }\equiv2\pi^{1/2}c_{s0}^{\left(  d\right)  }
\left[  \Gamma\left(  \frac{d}{2}+1\right)  \rho_{d}\right]  ^{\frac{1}{d}}\,.
\label{coeficiente Cd}%
\end{align}
In Eq.~\eqref{coeficiente Cd}, $\rho_{d}$ denotes the (hyper)volumetric
density of the cube, defined as
\begin{equation}
\rho_{d}\equiv\frac{n}{L^{d}}\,.
\end{equation}

The most physically relevant scenario is the three-dimensional solid. For
$d=3$, an exact solution for Eq.~\eqref{polinomial equation} can be obtained:
\begin{equation}
\omega_{D\pm}^{\left(  3\right)  }=\pi c_{s0}^{\left(  3\right)  }\left[
\sqrt[3]{f_{0\pm}}+\sqrt[3]{f_{1\pm}}\mp\frac{3\left(  c_{s0}^{\left(
3\right)  }\right)  ^{2}}{4\left(  c_{s\pm}^{\left(  3\right)  }\right)  ^{2}%
}\frac{1}{L}\right]  \,, \label{FreqDebye3D}%
\end{equation}
where
\begin{equation}
f_{p\pm}\equiv\frac{6\rho_{3}}{\pi}\mp\frac{27\left(  c_{s0}^{\left(
3\right)  }\right)  ^{6}}{32\left(  c_{s\pm}^{\left(  3\right)  }\right)
^{6}}\frac{1}{L^{3}}+\left(  -1\right)  ^{p}\left[  \frac{9\rho_{3}^{2}}%
{\pi^{2}}\pm\frac{81\left(  c_{0}^{\left(  3\right)  }\right)  ^{6}}{32\pi
^{2}\left(  c_{s\pm}^{\left(  3\right)  }\right)  ^{6}}\frac{\rho_{3}}{L^{3}%
}\right]  ^{1/2}\,. \label{FreqDebye3D-2}%
\end{equation}

Let us consider other scenarios besides the most usual one. In the
two-dimensional case, again an exact expression for the Debye frequency
$\omega_{D\pm}^{(2)}$ can be produced,
\begin{equation}
2 \omega_{D\pm}^{(2)} = \sqrt{B_{2\pm}^{2} + \left(  2 \omega_{D0}^{(2)}
\right)  ^{2} } - B_{2\pm} \, . \label{FreqDebye2D}%
\end{equation}
For $d > 3$, we use the approximation
\begin{equation}
\left(  \omega_{D\pm}^{\left(  d\right)  }\right)  ^{d}+B_{d\pm}\left(
\omega_{D\pm}^{\left(  d\right)  }\right)  ^{d-1}=\left(  \omega_{D\pm
}^{\left(  d\right)  }+\frac{B_{d\pm}}{d}\right)  ^{d}+O\left(  \frac{1}%
{L^{2}}\right)  \,,
\end{equation}
which is justified since we are considering the quasithermodynamic regime of
the theory. Therefore, keeping only terms of order $1/L$ and employing
result~\eqref{polinomial equation}, we obtain
\begin{equation}
\omega_{D\pm}^{\left(  d\right)  }=\omega_{D0}^{\left(  d\right)  }\mp
\frac{\pi^{1/2}\Gamma\left(  \frac{d}{2}+1\right)  }{\Gamma\left(  \frac
{d-1}{2}+1\right)  }\frac{\left(  c_{s0}^{\left(  d\right)  }\right)  ^{d}%
}{\left(  c_{s\pm}^{\left(  d\right)  }\right)  ^{d-1}}\frac{1}{L}\,.
\label{frequencia de debye dd}%
\end{equation}
It should be remarked that, for the two- and three-dimensional cases, exact
expressions~\eqref{FreqDebye2D} and \eqref{FreqDebye3D} are available. With
these results, higher-order contributions in $1/L$ are considered [when
compared to Eq.~\eqref{frequencia de debye dd}].

In the strict thermodynamic limit~\eqref{thermodynamic_limit} we observe that
$B_{d\pm}\rightarrow0$, and
\begin{equation}
\omega_{D\pm}^{\left(  d\right)  }\left(  L\rightarrow\infty\right)
=\omega_{D0}^{\left(  d\right)  }\,. \label{thermolimit-D}%
\end{equation}
\newline With $d=3$, expression~\eqref{thermolimit-D} reduces to the
well-known Debye frequency. Summarizing, we have obtained in
Eq.~\eqref{frequencia de
debye dd} a correction in Debye frequency for a harmonic solid in $d$
dimensions, with free ($+$) or fixed ($-$) walls.

It is interesting to notice that result~\eqref{frequencia de debye dd} could
offer a possible experimental test for the developments presented in this
work. Indeed, from Eq.~\eqref{frequencia de debye dd} we observe that
\begin{equation}
\omega_{D+}^{\left(  d\right)  }\leq\omega_{D0}^{\left(  d\right)  }\leq
\omega_{D-}^{\left(  d\right)  }\,. \label{desigualdades}%
\end{equation}
That is, a solid with free walls should have lower Debye frequency when
compared with the standard result in the strict thermodynamic regime. In a
similar way, a solid with fixed walls will have larger Debye frequency. These
differences are, in principle, measurable. Given that in an electric conductor
material the main contribution for the thermal capacity comes from the
electrons that are not strongly bound to the lattice, we expect that the
effect~\eqref{desigualdades} will be more relevant in an electric insulator
(non-metallic crystal).

\subsection{Heat capacity}

We now consider the heat capacity of the solid in the quasithermodynamic
regime. To evaluate the heat capacity at constant volume $C_{\pm}^{\left(
d\right)  }$, let us examine the internal energy $U_{\pm}^{\left(  d\right)
}$ of the system for both boundary conditions studied. Using Eq.~\eqref{Nsom},
we obtain that the density of modes is given by
\begin{equation}
D_{s\pm}^{\left(  d\right)  }=\frac{dN_{s\pm}^{\left(  d\right)  }}{d\omega
}=\kappa_{0}^{\left(  d\right)  }\omega^{d-1}+\kappa_{\pm}^{\left(  d\right)
}\omega^{d-2}\,, \label{DegeSom}%
\end{equation}
where the following quantities are defined:
\begin{align}
\kappa_{0}^{\left(  d\right)  }  &  \equiv\frac{d^{2}}{2^{d}\pi^{d/2}}%
\frac{L^{d}}{\Gamma\left(  \frac{d}{2}+1\right)  \left(  c_{s0}^{\left(
d\right)  }\right)  ^{d}}\,,\\
\kappa_{\pm}^{\left(  d\right)  }  &  \equiv\pm\frac{d^{2}\left(  d-1\right)
}{2^{d}\pi^{\frac{d-1}{2}}}\frac{L^{d-1}}{\Gamma\left(  \frac{d-1}%
{2}+1\right)  }\left(  c_{s\pm}^{\left(  d\right)  }\right)  ^{1-d}\,.
\label{kappaMAISMENOS}%
\end{align}
Hence, the internal energy can be written as
\begin{equation}
U_{\pm}^{\left(  d\right)  }=\frac{k_{B}^{2}T^{2}}{\hbar}\int_{0}^{\theta
_{\pm}^{\left(  d\right)  }/T}D_{s\pm}^{\left(  d\right)  }\left(  \frac
{k_{B}T}{\hbar} x \, \right)  \frac{x \, \mathrm{d}x}{ e^{x} - 1 } \,,
\label{ud}%
\end{equation}
with
\begin{equation}
\theta_{\pm}^{\left(  d\right)  }\equiv\frac{\hbar}{k_{B}}\omega_{D\pm
}^{\left(  d\right)  } \label{Temp de Debye}%
\end{equation}
denoting the Debye temperature in $d$ dimensions. In general, the integral in
Eq.~\eqref{ud} does not have an analytic solution. Let us investigate the
regimes of low temperature ($T\ll\theta_{\pm}^{\left(  d\right)  }$) and
high temperature ($T\gg\theta_{\pm}^{\left(  d\right)  }$).

The low-temperature limit is more commonly treated in the pertinent
literature. In three dimensions, this regime is captured by Debye law: the
specific heat of a solid at constant volume varies as the cube of the absolute
temperature $T$. We aim to improve Debye law in $d$ dimensions considering
solids with finite size, and hence adopting a quasithermodynamic description.

It should be remarked that the quasithermodynamic
limit~\eqref{quasithermodynamic_limit} implies $[k_{B}/(\hbar c)]\,LT\gg1$.
Therefore, low values for the temperature $T$ should be compensated by
corresponding large values for $L$. For low enough temperatures, the internal
energy of the system can be written as
\begin{equation}
U_{\pm}^{\left(  d\right)  }\left(  T\ll\theta_{\pm}^{\left(  d\right)
}\right)  =\frac{d!\kappa_{0}^{\left(  d\right)  }}{\hbar^{d}}\zeta\left(
d+1\right)  k_{B}^{d+1}T^{d+1}+\frac{\left(  d-1\right)  !\kappa_{\pm
}^{\left(  d\right)  }}{\hbar^{d-1}}\zeta\left(  d\right)  k_{B}^{d}T^{d}\,.
\label{exp-U}%
\end{equation}
From expression~\eqref{exp-U} for $U_{\pm}^{\left(  d\right)  }$, corrections
in Debye law can be obtained:
\begin{equation}
C_{\pm}^{\left(  d\right)  }\left(  T\ll\theta_{\pm}^{\left(  d\right)
}\right)  =\frac{d!\left(  d+1\right)  \kappa_{0}^{\left(  d\right)  }}%
{\hbar^{d}}\zeta\left(  d+1\right)  k_{B}^{d+1}T^{d}+\frac{d\left(
d-1\right)  !\kappa_{\pm}^{\left(  d\right)  }}{\hbar^{d-1}}\zeta\left(
d\right)  k_{B}^{d}T^{d-1}\,. \label{DebyeCORR}%
\end{equation}
In the strict thermodynamic limit~\eqref{thermodynamic_limit} we recover the
Debye law in $d$ dimensions \cite{valladares}. For the three-dimensional case,
expression~\eqref{DebyeCORR} reduces to the result previously discussed in
\cite{Dupuis}.

The high-temperature regime is less explored, and we will consider it in the
present work. This scenario is actually more suitable for the
quasithermodynamic treatment, because if $T$ is high, solids with low size $L$
can be more accurately analyzed. In three dimensions, assuming the
thermodynamic limit, the behavior of the system in this regime is captured by
Dulong-Petit law: the heat capacity of a solid with a mol of particles is
approximated constant for high enough temperature. Corrections for the
Dulong-Petit law in $d$ dimensions considering a finite-size solid will be derived.

In the high-temperature regime $\theta_{\pm}^{\left(  d\right)  }%
/T\rightarrow0$, and the approximation $e^{x}-1\approx x$ can be employed in
Eq.~\eqref{ud}. With this approach, the internal energy $U_{\pm}^{\left(
d\right)  }(T\gg\theta_{\pm}^{\left(  d\right)  })$ can be explicitly written
and the thermal capacity $C_{\pm}^{\left(  d\right)  }$ determined:
\begin{equation}
C_{\pm}^{\left(  d\right)  }\left(  T\gg\theta_{\pm}^{\left(  d\right)
}\right)  =\left(  \frac{k_{B}}{\hbar}\right)  ^{d}\left[  \frac{\kappa
_{0}^{\left(  d\right)  }}{d}\left(  \theta_{\pm}^{\left(  d\right)  }\right)
^{d}+\frac{\hbar\kappa_{\pm}^{\left(  d\right)  }}{k_{B}\left(  d-1\right)
}\left(  \theta_{\pm}^{\left(  d\right)  }\right)  ^{d-1}\right]  k_{B}\,.
\label{DulongPetitiCORR}%
\end{equation}
In the strict thermodynamic limit~\eqref{thermodynamic_limit} we observe that
$\omega_{D\pm}\rightarrow\omega_{D0}^{\left(  d\right)  }$ and hence
\begin{equation}
C_{\pm}^{\left(  d\right)  }\left(  T\gg\theta_{\pm}^{\left(  d\right)
},\text{ }L\rightarrow\infty\right)  =\frac{\kappa_{0}^{\left(  d\right)
}\omega_{0}^{\left(  d\right)  }}{d}k_{B}\,. \label{DulongPetitiDd}%
\end{equation}
For the three-dimensional case, result~\eqref{DulongPetitiDd} is reduced to
the usual Dulong-Petit law. For general values of $d$,
Eq.~\eqref{DulongPetitiCORR} furnishes the correction of the Dulong-Petit law
for a finite solid of arbitrary dimensionality.

\section{\label{conclusion} Final comments}

In the present work, we explored Weyl law and its extensions with an intuitive
formalism, based on the association between point counting and volumes of
sections of the sphere. In several published developments, the counting
function depended strongly on the pertinent domain, with the associated
lattice kept fixed. In our approach, the domain is kept fixed and the lattice
is rescaled. Known results were rederived, showing the robustness of the
method. Moreover, new results were obtained, including corrections for the
Weyl conjecture in $d$ dimensions, effects of the polarization, and an
exploration of the role of the area term in the three-dimensional scenario.
Applications of the previous results were investigated, with the
quasithermodynamic analyses of the electromagnetic field in a finite cavity
and acoustic perturbations in a finite solid.

Applying the formalism to the thermodynamics and quasithermodynamics of the
electromagnetic perturbations in a finite box within a semi-classical treatment, corrections to the
$d$-dimensional Stefan-Boltzmann law were obtained and polarization and border
effects were treated. In particular, we showed that the well-known
cancellation of the area term only occurs in three dimensions. This effect
turns the thermodynamics of the system distinct for $d\neq3$. The correction
due to a minimum energy of the system is treated. In all scenarios except the
three-dimensional case, the quasithermodynamic corrections suppress an
eventual effect of a minimal energy.

Two-dimensional results for the quasithermodynamics of the electromagnetic
field can be linked to experimental setups. Indeed, the analysis of thermal
radiation in $d=2$ has applications in the description of single-layer
materials (also known as \textquotedblleft2D materials\textquotedblright),
which include graphene, single layers of various dichalcogenides and complex
oxides \cite{A15}. The results presented in \cite{stefan 2d}, describing
two-dimensional thermal radiation and improved in the present work, were used
in \cite{B1} to study the emission spectra of a graphene transistor.

Concerning the electromagnetic field in a three-dimensional cavity, there is
some discussion in the literature involving the cancellation of area term. We
believe this controversy is the result of an inadequate treatment of how
polarization affects each term of the expansion~\eqref{eQ39}. For instance, in
\cite{Maslov1} polarization is assumed to have the same effect on all terms of
the expansion~\eqref{eQ39}, and as a result the cancellation of the area term
does not occur. Our result in Eq.~\eqref{nk0} indicates that this approach is
inadequate. Experimentally, the cancellation of the area term in a
three-dimensional setup implies in a very small correction \cite{Baltes
Baltes}. Therefore, other effects (such as scattering and diffraction) can
supplant this correction. In fact, while the results in \cite{Quinn} are
compatible with the negative value correction presented in
Eq.~\eqref{N-elect-d3}, other experiments point to a positive correction term
\cite{Datla}. 

Moreover, we believe that three-dimensional systems are not the most suitable
setups for the observation of effects associated to the Weyl conjecture on
electromagnetic radiation. Our results suggest that a better approach to this goal is the use of two-dimensional effective systems. Such systems could be constructed using single-layer or nanostructures graphene devices simulating two-dimensional blackbodies \cite{B1,Takahiro}.

Another main application of the developed formalism concerns the
thermodynamics of acoustic perturbations. An improved version of Debye model for a finite
solid is treated. We reproduce the known results for the three-dimensional
case in the limit of low temperatures and extend those results to arbitrary
dimensions. New developments in the high-temperature scenario and the
influence of the area term are explored. Extensions of the known formulas are
obtained for Debye temperature and Dulong-Petit law.

The presented development captures some effects associated to internal degrees of freedom, such as spin. Indeed, macroscopic effects associated to spin manifest itself in the polarization of the fields. This is more directly seen when the electromagnetic perturbation is considered. In this case, the two values of the photon spin (or, more precisely, of the photon helicity) can be related to the two polarizations of the classical field. For acoustic perturbations the problem is more subtle~\cite{Levine}. In an ideal isotropic medium, considering modes with long wavelength, it is possible to associate longitudinally polarized modes with spin-0 phonons, and transversely polarized modes with spin-1 phonons. 
In general, new internal degrees of freedom can be incorporated into the presented approach, as long as they do not modify the eigenvalue problem [i.e., the Helmholtz equation~\eqref{helmholtz d dimensoes}]. This can be done by increasing the multiplicity count of the modes performed in section~\ref{comment}.

\newpage

The analysis of arbitrary boundary conditions in the extensions of Weyl law,
related with all self-adjoint extensions of the $d$-dimensional Laplacian
operator, is a work in progress. In fact, as far as we know, there is no Weyl
conjecture for such general scenario. We believe that the approach introduced
in the present development, based on a counting function with the rescaling of
the lattice, might be an important step in the problem. Further developments
of the present work might include the application of the proposed formalism on
gravitational systems. Also, a possible exploration of the influence of the
area term in the Casimir effect could be conducted, considering
three-dimensional systems with finite temperature. Analyses along those lines
should appear in forthcoming presentations.

\appendix

\section{\label{eixos} Hypervolume associated to the axes}

The calculation of the hypervolume associated to the axes in the
$d$-dimensional case will be presented. Before the actual calculation, let us
fix the notation and definitions that are used in the development.

We denote $\tilde{\Omega}_{d}$ as one of the $2^{d}$ partitions of the
$d$-dimensional unit sphere, that is,
\begin{equation}
\tilde{\Omega}_{d}\equiv\left\{  (\epsilon_{1},\epsilon_{2},\ldots
,\epsilon_{d})\in\mathbb{R}^{d}:0\leqslant\epsilon_{i}\leqslant1\ \text{and}%
\ \epsilon_{1}^{2}+\epsilon_{2}^{2}+\cdots+\epsilon_{d}^{2}\leqslant1\right\}
\,. \label{OMEGAD}%
\end{equation}
Also, the hyperplanes $\mathcal{H}_{i}$ are defined as
\begin{equation}
\mathcal{H}_{i}\equiv\left\{  (\epsilon_{1},\epsilon_{2},\ldots,\epsilon
_{d})\in\mathbb{R}^{d}:\epsilon_{i}=0\right\}  \, , \,\, i=1,2,\ldots,d\,.
\label{hiperplanos}%
\end{equation}
The volume of a given subset $S\subset\mathbb{R}^{d}$ is indicated by $|S|$.
For example,
\begin{equation}
\left\vert \tilde{\Omega}_{d}\right\vert =2^{-d}\omega_{d}\,. \label{Volomega}%
\end{equation}

Using the sets $\tilde{\Omega}_{d}$ and $\mathcal{H}_{i}$ defined in
Eqs.~\eqref{OMEGAD} and \eqref{hiperplanos}, the subsets $\{\Pi_{i_{1}%
,\ldots,i_{n}}\}$ of $\tilde{\Omega}_{d}$ are constructed:
\begin{equation}
\Pi_{i_{1},\ldots,i_{n}}=\tilde{\Omega}_{d}\cap\mathcal{H}_{i_{1}}\cap
\cdots\cap\mathcal{H}_{i_{n}} \, , \,\, 1\leqslant i_{n}\leqslant d \, , \,\, 1\leqslant
n\leqslant d\,. \label{Fdef}%
\end{equation}
The number of subsets $\{\Pi_{i_{1},\ldots,i_{n}}\}$ is $n^{d}$. However, some
of this subsets are equal, because: (1) $\Pi_{i_{1},\ldots,i_{n}}$ is
symmetric under the switch of indexes; (2) and the idempotent property
$\mathcal{H}_{j}\cap\mathcal{H}_{i}=\mathcal{H}_{i}$, for $i=j$. So, the
number of distinct sets is
\begin{equation}
\#\Pi_{i_{1},\ldots,i_{n}}=\binom{d}{n}=\frac{d!}{n!\left(  d-n\right)  !}\,.
\end{equation}

The subsets $\{\Pi_{i_{1},\ldots,i_{n}}\}$, each one labeled by $n$ indexes,
can be interpreted as sections of the sphere in $\mathbb{R}^{d-n}$, generated
by different axes. For example, if $d=3$,
\begin{equation}
(n=1)
\begin{cases}
\Pi_{1}=\Omega_{2}\ \text{generated by}\ \epsilon_{2}\ \text{and}%
\ \epsilon_{3}\\
\Pi_{2}=\Omega_{2}\ \text{generated by}\ \epsilon_{1}\ \text{and}%
\ \epsilon_{3}\\
\Pi_{3}=\Omega_{2}\ \text{generated by}\ \epsilon_{1}\ \text{and}%
\ \epsilon_{2}%
\end{cases}
, \, (n=2)%
\begin{cases}
\Pi_{12}=\Omega_{1}\ \text{generated by}\ \epsilon_{3}\\
\Pi_{13}=\Omega_{1}\ \text{generated by}\ \epsilon_{2}\\
\Pi_{23}=\Omega_{1}\ \text{generated by}\ \epsilon_{1}%
\end{cases}
, \, (n=3) \left\{  \Pi_{123}=\left(  0,0,0\right)  \right.  \,.
\end{equation}
With the symbology $\Pi_{i_{1},\ldots,i_{n}}$, $n$ indicates the number of
axes not included in the subset generation.

From a given section $\Pi_{i_{1},\ldots,i_{n}}$, a cylinder $\mathcal{C}^{I}$
can be defined, adding to $\Pi_{i_{1},\ldots,i_{n}}$ an interval with the form
$I=[a,b]$ for each not included axes,
\begin{equation}
\mathcal{C}_{i_{1},\ldots,i_{n}}^{I}=\Pi_{i_{1},\ldots,i_{n}}\times
\underbrace{I\times I\times\cdots\times I}_{n}\,.
\end{equation}
It should be observed that, for each value of $n$, there are $\binom{d}{n}$
different cylinders with the same volume $V_{d}^{n}$, where
\begin{equation}
V_{d}^{n} \equiv\left\vert \mathcal{C}_{i_{1},\ldots,i_{n}}^{I}\right\vert
=|I|^{n}\left\vert \Pi_{i_{1}, \ldots,i_{n}}\right\vert =|I|^{n}2^{n-d}%
\omega_{d-n} \, .
\end{equation}

Using previous results, the contribution of the axes to the total volume can
be determined. Considering the volumes of the cylinders $\mathcal{C}^{I}$
constructed with the interval $I=\left[  -\frac{\epsilon}{2},0\right]  \,$, we
obtain
\begin{equation}
\text{Hypervolume of the axes}=\sum_{n=1}^{d}{\binom{d}{n}V_{d}^{n}}%
=\sum_{n=1}^{d}{\binom{d}{n}\left(  \frac{\epsilon}{2}\right)  ^{n}%
2^{n-d}\omega_{d-n}}=\sum_{n=1}^{d}{\binom{d}{n}2^{-d}\omega_{d-n}\epsilon
^{n}\,.} \label{PARTIALOMEGA}%
\end{equation}

\section{\label{pascal}Counting functions}

Let us see how Neumann counting function can be constructed from the counting
function for Dirichlet case and vice versa. We start by arranging the Neumann
modes in $d+1$ classes, labeled by $j=0,1,\ldots,d$. The first class is
composed by modes where all $n_{i}$ are non-null. The second class is composed
by modes where only one of the $n_{i}$ is zero. In the third class two of the
$n_{i}$ are zero. The process is continued in this fashion, up to $d+1$ sets.

Given the proposed partition of the Neumann modes, we observe that the class
where none of the $n_{i}$ is zero is the solution for the Dirichlet problem in
$d$ dimensions, where all modes with $n_{i}=0$ must be excluded. The class
with only one of the $n_{i}$ is zero can be further divided into
$d$\ sub-classes where $n_{1}=0$, or $n_{2}=0$, etc. Hence, in each one of
these sub-classes, disregarding the null $n_{i}$, we obtain sub-classes
composed by Dirichlet solutions in $d-1$ dimensions. The class where two
$n_{i}$ are null can be separated into $\binom{d}{2}$ sub-classes composed of
modes which satisfy Dirichlet boundary conditions in $d-2$ dimensions.
Carrying on with this procedure, solutions of the Neumann problem can be
written as a combination of the elements in the constructed sub-classes:
\begin{equation}
N_{+}^{\left(  d\right)  }\left(  \epsilon\right)  =\sum_{j=0}^{d}{\binom
{d}{j}N_{-}^{\left(  d-j\right)  }}\left(  \epsilon\right)  \,. \label{nm}%
\end{equation}

On the other hand, we know that the solutions for Dirichlet, unlike those for
Neumann, exclude the mode $n_{i}=0$, and therefore the difference between the
number of solutions of both will be given by
\begin{equation}
N_{+}^{\left(  d\right)  }-N_{-}^{\left(  d\right)  }=\sum_{n}\text{(\#
Dirichlet solutions formed by $n$ different axes)}\,.
\end{equation}
Since there are $d$ axes, for each set of $n$ axes there will be a degeneracy
of $\binom{d}{n}$ in the Dirichlet solutions, and hence
\begin{equation}
N_{+}^{\left(  d\right)  }-N_{-}^{\left(  d\right)  }=\sum_{n=0}^{d-1}%
\binom{d}{n}N_{-}^{\left(  n\right)  }\,.\label{eqquase}%
\end{equation}
Substituting $N_{-}^{(d)}$ in the sum~\eqref{eqquase}, we get the relation~\eqref{nm}.

The inverse relation is obtained writing Eq.~\eqref{nm} as
\begin{equation}
N_{+}^{\left(  i\right)  }=\sum_{j=0}^{d}P_{j}^{i}N_{-}^{\left(  j\right)
}\,,\ P_{j}^{i}=\left\{
\begin{array}
[c]{l}%
\binom{i}{j},\ \ j\leq i\\
0\,\ \ \ \ j>i
\end{array}
\right.  \,,\ i,j=0,1,2,...,d~.\label{PascalM}%
\end{equation}
The triangular matrix $P$ is known as the Pascal's matrix. It is invertible,
since $\det P=1$, with an inverse given by
\begin{equation}
\left(  P^{-1}\right)  _{j}^{i}=\left\{
\begin{array}
[c]{l}%
\left(  -1\right)  ^{i-j}\binom{i}{j},\ \ j\leq i\\
0,\ \ j>i
\end{array}
\right.  \,.
\end{equation}
Therefore we can solve Eq.~\eqref{PascalM} for $N_{-}^{\left(  d\right)  }$,
obtaining
\begin{equation}
N_{-}^{\left(  d\right)  }=\sum_{n=0}^{d}\left(  -1\right)  ^{d-n}\binom{d}%
{n}N_{+}^{\left(  n\right)  }.\label{nm2}%
\end{equation}

\begin{acknowledgments}
L.~F.~S. acknowledges the support of Coordena\c{c}\~{a}o de Aperfei\c{c}oamento de Pessoal
de N\'{\i}vel Superior (CAPES) -- Brazil, Finance Code 001; and the Funda\c{c}\~{a}o
Arauc\'{a}ria (Foundation in Support of the Scientific and Technological
Development of the State of Paran\'{a}, Brazil). M.~A.~J. thanks Coordena\c{c}\~{a}o de
Aperfei\c{c}oamento de Pessoal de N\'{\i}vel Superior (CAPES) -- Brazil, Finance Code
001, for financial support.  
This is the version of the article before peer review or editing, as submitted by to Journal of Physics A. IOP Publishing Ltd is not responsible for any errors or omissions in this version of the manuscript or any version derived from it. The Version of Record is available online at \url{https://doi.org/10.1088/1751-8121/acb09b}.
\end{acknowledgments}

\end{document}